\documentclass[floatfix,aps,unsortedaddress,superscriptaddress,showpacs]{revtex4}

\usepackage[dvips]{graphicx}
\usepackage{epsfig}
\begin{document}

\title{Indirect techniques in nuclear astrophysics. Asymptotic Normalization Coefficient and Trojan Horse}

\author{A.\,M.~Mukhamedzhanov}
\affiliation{Cyclotron Institute, Texas A\&M University,
College Station, TX 77843}

\author{L. D. Blokhintsev}
\affiliation{Institute of Nuclear Physics, Moscow State University, Moscow, Russia}

\author{B. A. Brown}
\affiliation{N.S.C.L. and Department of Physics and Astronomy, 
Michigan State University, East Lansing, MI, USA} 

\author{V. Burjan}
\affiliation{Nuclear Physics Institute of Czech Academy of Sciences, 
Prague-{\v R}e{\v z},\,Czech Republic}

\author{S. Cherubini}
\affiliation{DMFCI, Università di Catania, Catania, Italy and INFN - Laboratori Nazionali del Sud, Catania, Italy}

\author{C. A. Gagliardi}
\affiliation{Cyclotron Institute, Texas A\&M University,
College Station, TX 77843}

\author{B. F. Irgaziev}
\affiliation{Physics Department, National University, Tashkent, Uzbekistan }
 
\author{V. Kroha}
\affiliation{Nuclear Physics Institute of Czech Academy of Sciences, 
Prague-{\v R}e{\v z},\,Czech Republic}

\author{F. M. Nunes}
\affiliation{N.S.C.L. and Department of Physics and Astronomy, 
Michigan State University, East Lansing, MI, USA}

\author{F. Pirlepesov}
\affiliation{Cyclotron Institute, Texas A\&M University,
College Station, TX 77843}

\author{R. G. Pizzone}
\affiliation{DMFCI, Università di Catania, Catania, Italy and INFN - Laboratori Nazionali del Sud, Catania, Italy}

\author{S. Romano}
\affiliation{DMFCI, Università di Catania, Catania, Italy and INFN - Laboratori Nazionali del Sud, Catania, Italy}

\author{C. Spitaleri}
\affiliation{DMFCI, Università di Catania, Catania, Italy and INFN - Laboratori Nazionali del Sud, Catania, Italy}

\author{X. D. Tang}
\affiliation{Physics Division, Argonne National Laboratory, Argonne, Illinois 60439, USA}

\author{L. Trache}
\affiliation{Cyclotron Institute, Texas A\&M University,
College Station, TX 77843}

\author{R.E. Tribble}
\affiliation{Cyclotron Institute, Texas A\&M University,
College Station, TX 77843}

\author{A. Tumino}
\affiliation{DMFCI, Università di Catania, Catania, Italy and INFN - Laboratori Nazionali del Sud, Catania, Italy}

\begin{abstract}Owing to the presence of the Coulomb barrier at 
astrophysically relevant kinetic energies it is very difficult, or
sometimes impossible, to measure astrophysical reaction rates in the
laboratory. That is why different indirect techniques
are being used along with direct measurements. Here we address
two important indirect techniques, the asymptotic normalization
coefficient (ANC) and the Trojan Horse (TH) methods. We discuss
the application of the ANC technique for calculation of the
astrophysical processes in the presence of subthreshold bound
states, in particular, two different mechanisms are discussed:
direct capture to the subthreshold state and capture to the
low-lying bound states through the subthreshold state, which plays
the role of the subthreshold resonance. The ANC technique can also
be used to determine the interference sign of the resonant and
nonresonant (direct) terms of the reaction amplitude. The TH method is 
unique indirect technique allowing one to measure astrophysical
rearrangement reactions down to astrophysically relevant energies.
We explain why there is no Coulomb barrier in the sub-process
amplitudes extracted from the TH reaction.  The expressions for 
the TH amplitude for direct and resonant cases are presented.
\end{abstract} 
\pacs{26.20.+f, 21.10.Jx, 25.55.Hp, 27.20.+n}

\maketitle

\section{Introduction}
\label{intro}
For better understanding stellar evolution, cross sections of 
astrophysically relevant nuclear reactions should be known at the
Gamow energy with an accuracy better than $10\%$ \cite{rolfs88}.
The presence of the Coulomb barrier for colliding charged nuclei
makes nuclear reaction cross sections at astrophysical energies so
small that their direct measurements in laboratories is very
difficult, or even impossible. That is why direct measurements are being 
done at higher energies and then extrapolated down to the Gamow energy. 
Such an extrapolation procedure can cause an additional uncertainty. 
Also for nuclear reactions studied in laboratory, the electron clouds 
surrounding the interacting nuclei lead to a screened cross section which is
larger than the "bare" nucleus one (see
\cite{ass87,streider01,cas02,spit04} and references therein). The
enhancement factor is determined by the electron screening
potential  which is a model dependent quantity and its value in
the laboratory is different from the one present in the stellar
environment.  There are four often used indirect
techniques: the Asymptotic normalization coefficient (ANC) method 
\cite{muk03}, Coulomb breakup processes \cite{baur86,mot94}, Trojan Horse (TH) 
\cite{baur86th,spit04} and the Surrogate reactions 
method (see \cite{younes03} and references therein). In this work we address only two indirect techniques, the ANC and TH methods.

\section{ANC method}
\label{sec:1} The ANC method has been suggested in
\cite{mt90,xu94} and can be used to determine the astrophysical
factors for peripheral radiative capture processes. The method can
be applied for analysis of direct radiative capture processes
leading to final loosely bound states. Due to small binding
energies and strong Coulomb barrier, the direct capture reactions
are peripheral. In previous papers \cite{mt90,xu94,gag99} it has
been pointed out that the overall normalization of the cross
section for a direct radiative capture reaction at low binding
energy is entirely defined by the ANC of the final bound state
wave function into the two-body channel corresponding to the
colliding particles. The ANC technique turns out to be very
productive for analysis of the astrophysical processes in the
presence of the subthreshold state \cite{mukhtr99}. Here we
address some applications of the ANC method in the presence of the
subthreshold state. We also demonstrate how ANC technique can be
used to determine the interference sign of the direct and resonant
amplitudes for some important astrophysical radiative capture
reactions.
\subsection{Definition of the ANC}
\label{defANC1}
We present first some useful equations for the ANC. Let us
consider a virtual decay of nucleus c into two nuclei a and b.
First we introduce the overlap function $I$ of the bound state
wave functions of particles $c$, $a$, and $b$ \cite{blokh77} :
\begin{eqnarray}
I_{ab}^c ({\rm {\bf r}})&=& <
\varphi_a(\zeta_a)\,\varphi_b(\zeta_b)|\varphi_c(\zeta_a ,\zeta_b ;{\rm
{\bf r}})>  \nonumber\\
&=& \sum\limits_{l_c m_{l_c } j_c m_{j_c}}i^{l_c }< J_a M_a j_c m_{j_c }|
J_c M_c>  \nonumber\\
&\times&< J_b M_b \,l_c m_{l_c }|j_c m_{j_c}  >  Y_{l_c
m_{l_c}}({\hat{\rm {\bf r}}})\,I_{abl_c j_c }^c (r).
 \label{ovrlint1}
\end{eqnarray}
Here $\varphi_{i},\,\zeta_{i},\, J_{i}$ and $\,M_{i}$ are the bound state wave function, a set of internal coordinates including spin-isospin variables, spin and spin projection for nucleus $i$. Also r is the relative coordinate of the centers of
mass of nuclei a and b, ${\hat{\rm {\bf r}}}={\rm {\bf r}}/r$,
$\,j_{c},\, m_{j_{c}}$ are the total angular momentum of particle
b and its projection in the nucleus $c=(ab)$, $\,l_{c},\,
m_{l_{c}}$ are the orbital angular momentum of the relative motion
of particles $a$ and $b$ in the bound state $c=(ab)$ and its
projection, $<j_{1}m_{1}j_{2}m_{2}|j_{3}m_{3}$ is a Clebsch-Gordan
coefficient, $Y_{l_{c}m_{c}}({\hat{\rm {\bf r}}})$ is a spherical
harmonic, and $I_{abl_c j_c }^c (r)$ is the radial overlap
function which includes the antisymmetrization factor due to
identical nucleons. The summation over $l_{c}$ and $j_{c}$ is
carried out over the values allowed by angular momentum and parity
conservation in the virtual process $c \to a+b$. The asymptotic
normalization coefficient $C_{abl_c j_c }^c$ defining the
amplitude of the tail of the radial overlap function $I_{abl_c j_c
}^c (r)$ is given by  \cite{blokh77}
\begin{equation}
I_{abl_c j_c }^c (r) \stackrel{r >R_{N}}{\longrightarrow} C_{abl_c j_c 
}^c \frac{{W_{ - \eta _c ,l_c + 1/2} (2\kappa _{ab} r)}}{r},
\label{asympovrint1}
\end{equation}
where $R_{N}$ is the nuclear interaction radius between $a$ and $b$,
$W_{-\eta_c ,l_c + 1/2}(2\kappa_{ab}r)$ is the Whittaker function 
describing the asymptotic behavior of the bound state wave
function of two charged particles,
$\kappa=\sqrt{2\,\mu_{ab}\,\varepsilon_{c}}$ is the wave
number of the bound state $c=(ab)$, $\,\mu_{ab}$ is the reduced
mass of particles $a$ and $b$, $\varepsilon_{c}$ is the binding energy
of the bound state $(ab)$ and 
$\eta_{c}=Z_{a}\,Z_{b}\,e^{2}\,\mu_{ab}/\kappa$ is the Coulomb 
parameter of the bound state $(ab)$, $\,Z_{i}\,e$
is the charge of particle $i$. We use the system of units such that 
$\hbar=c=1$.
There is another definition of the ANC, the most model independent
one. The elastic $a+b$ scattering amplitude  in the channel
$(l_{c},j_{c})$ has a  pole in the momentum plane \cite{mukhtr99}
\begin{equation}
M_{l_{c}j_{c}}(k)= \frac{S_{l_{c}j_{c}} -1}{2\,i\,k} \stackrel{k \to 
k_{p}}{\longrightarrow}  \frac{1}{2\,i\,k_{p}}\,\frac{W_{l_{c}j_{c}}}{k 
- k_{p}}.
\label{ANCresidue1}
\end{equation}
corresponding to the bound state $c=(ab)$ for $k_{p}=i\,\kappa$ and 
to
the resonance for $k_{p}=k_{R}$, where $k_{R}=k_{0} - i\,k_{I}$ is 
the resonance location in the momentum plane. Here, $S_{l_{c}j_{c}}$ is 
the elastic matrix element of the $S$-matrix.
The residue in the pole $W_{l_{c}j_{c}}$ is
\begin{equation}
W_{l_{c}j_{c}}=-( - 1)^{l_c}\,ie^{i\pi \eta _c }\,{(C_{abl_c j_c 
}^c)}^2, \quad k_{p}=i\,\kappa,
\label{residbndst1}
\end{equation}
\begin{equation}
W_{l_{c}j_{c}}=-( - 1)^{l_c }\,i\,{(C_{abl_c j_c (R)}^c)}^2, \quad 
k_{p}=k_{R}.
\label{residbndst2}
\end{equation}
For narrow resonances, $k_{I} << k_{0}$,
\begin{equation}
{(C_{abl_c j_c (R)}^c)}^2= 
(-1)^{l_{c}}\,\frac{\mu_{ab}}{k_{1}}e^{\pi\,\eta_{0}}\,e^{2i\,\delta_{l_{
c}j_{c}}(k_{0})}\,\Gamma_{l_{c}j_{c}}.
\label{resANC1}
\end{equation}
Here $\eta_{0}$ is the Coulomb parameter for the resonance at
momentum $k_{0}$, $\,\delta_{l_{c}j_{c}}(k_{0})$ is the
potential (non-resonant) scattering phase shift taken at the
momentum $k_{0}$. Thus the residue in the bound state or resonance
pole is expressed in terms of the ANC and for the resonance the
ANC can be expressed in terms of the  partial resonance width
\cite{mukhtr99}.
Note that Eq. (\ref{ANCresidue1}) holds only for $k$ in the
closest vicinity of the pole. For elastic scattering at positive
energies in the presence of the Coulomb barrier, the elastic
scattering amplitude with the bound state pole behaves (in the $R$
matrix approach) as
\begin{equation}
M_{l_{c}j_{c}}(k)\stackrel{k \to 0}{\longrightarrow}
= -\frac{1}{2\,k}\,e^{ - 2i(\phi _{l_c}  - \sigma _{l_c } 
)}\,\frac{\Gamma_{c}}{E + \varepsilon _c  + i\,\Gamma_{c}/2}.
\label{ANCresidue2}
\end{equation}
where
\begin{equation}
\Gamma_{c}=2\,P_{l_c } (E)\,\gamma _c^2.
\label{reswidth1}
\end{equation}
Here $P_{l_c } (E)$ is the penetrability through the
Coulomb-centrifugal barrier, $\phi _{l_c}$ is the solid sphere 
scattering phase shift in the partial wave $l_{c}$ and
$\sigma_{l_{c}}=\sum\limits_{n=1}^{l_{c}}\,
\tan^{-1}(\frac{\eta_{c}}{n}) $ , $r_{0}$ is the channel radius,
$\gamma _c^2$ is the effective (observable) reduced width:
\begin{equation}
\gamma _c^2  = \frac{1}{{2\mu _{ab} }}\frac{{W_{ - \eta _c ,l_c  + 
1/2}^{} (2\kappa r_0 )}}{{r_0 }}\,{(C_{abl_c j_c (r)}^c)}^2. 
\label{redwidthANC1}
\end{equation}
Thus at positive energies, $E \to +0$ due to the presence of the
Coulomb-centrifugal barrier the elastic scattering amplitude
behaves as the resonant scattering amplitude with the resonance
width expressed in terms of the ANC. At positive energies the
elastic scattering cross section in the presence of the bound
state and the barrier behaves as the high-energy tail of the
resonance located at energy $E=-\varepsilon_{c}$. That what is
called the "subthreshold" resonance. However, it is
not a resonance because the real resonance is located at complex
energies on the second energy sheet, while the subthreshold
resonance is just the bound state located on the first energy
sheet at negative energy, corresponding to the bound state. At
negative energies (positive imaginary momenta) Eq.
(\ref{redwidthANC1}) reduces to Eq. (\ref{ANCresidue1}).
Definitions of the ANC dictate the experimental methods of its
determination. The ANC can be determined from peripheral transfer
reactions which are dominated by the tail of the overlap function.
Eq. (\ref{ANCresidue1}) offers another possibility to determine
the ANC, namely, by extrapolating the elastic scattering amplitude
(or equivalently the phase shift) to the bound state pole
\cite{blokh93}.
\subsection{ANC and astrophysical processes}
\label{ANCastroph1}
(i) For peripheral direct radiative capture reaction $a+b \to c+
\gamma$ to the final state $l_{c}j_{c}$ proceeding through the
$EL$ transition, the cross section is
\begin{eqnarray}
\sigma &\sim& |<I_{abl_c j_c }^c (r)|r^{L}|\psi_{k_{i}l_{i}}(r)>|^{2} 
\nonumber\\
&\approx& |C_{abl_c j_c }^c |^{2}|<\frac{{W_{ - \eta _c ,l_c + 1/2} 
(2\kappa _{ab} r)}}{r}|r^{L}|\psi_{k_{i}l_{i}}(r)>|^{2}. \nonumber\\
\label{dccrsct1}
\end{eqnarray}
Here $L$ is the multipolarity of transition,
$\psi_{k_{i}\,l_{i}}(r)$ is the initial $a+b$ scattering wave
function with the relative momentum $k_{i}$ in the partial wave
$l_{i}$. 
Thus the ANC determines the overall normalization
of the direct radiative capture cross sections. \\
(ii) The elastic scattering amplitude (\ref{ANCresidue2})
describes the elastic scattering through the intermediate bound state
$c=(ab)$. Assume that it is an excited state. Then, when the excited  
bound state is formed it can decay into the ground state by emitting
a photon. In this case we have the radiative capture 
process which is called the capture to the ground state through the 
subthreshold resonance. The amplitude of this process is given by
\begin{equation}
M_{l_{c}j_{c}}(k)\stackrel{k \to 0}{\longrightarrow}
= -\frac{1}{2\,k}\,e^{ - 2i(\phi _{l_c}  - \sigma _{l_c } 
)}\,\frac{\Gamma_{c}^{1/2}\,\Gamma_{\gamma}^{1/2}}{E + \varepsilon _c  + 
i\,\Gamma_{c}/2}.
\label{ANCresidue3}
\end{equation}
Here $|\Gamma_{\gamma}^{1/2}|^{2}$ gives the radiative width
for the transition from the excited bound state $\to $ ground state.
Thus in the presence of an excited bound state close to 
threshold, two different radiative capture processes can occur:
direct capture to this excited bound state or capture to the
low-lying bound states through this subthreshold bound state
(capture through the subthreshold resonance). In what follows we
present some astrophysical reactions in the presence of the
subthreshold state.
\subsection{ANC for ${}^{14}$N + $p$ $\to$ ${}^{15}$O and the 
astrophysical $S$ factor for ${}^{14}$N($p$,$\gamma$)${}^{15}$O}
\label{ANCastroph1} 
\begin{figure}[tbp]
\includegraphics*[width=\linewidth]{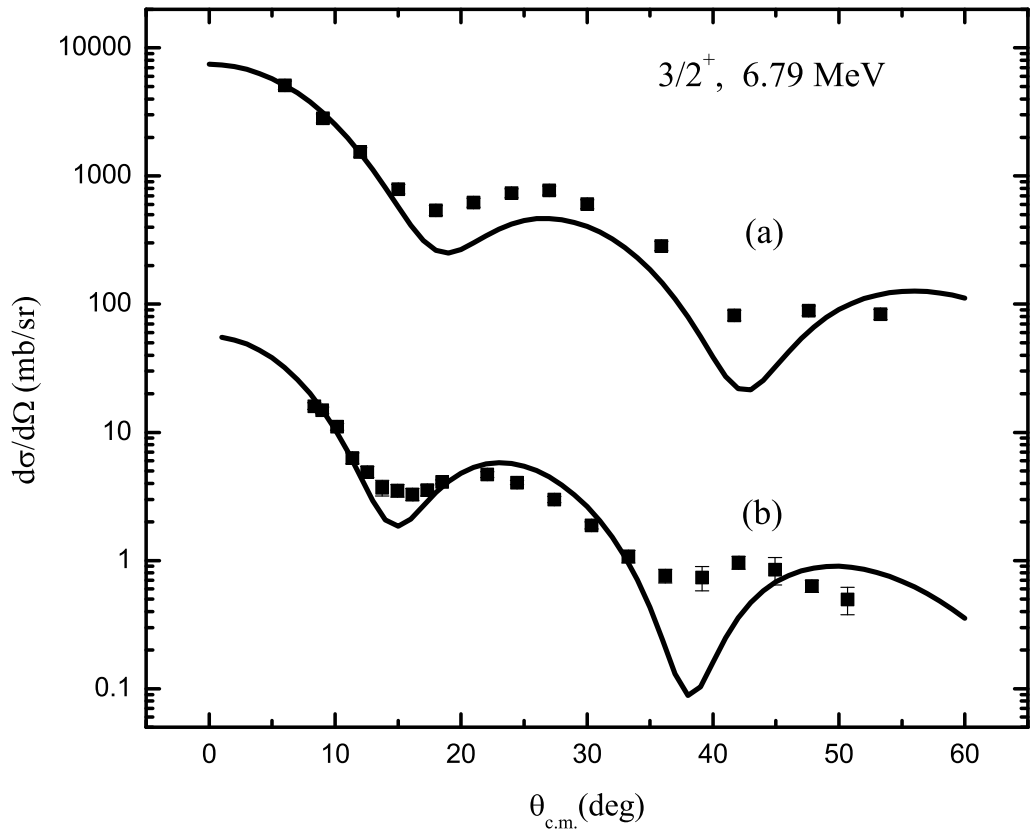}
\caption{The ${}^{14}{\rm N}({}^{3}{\rm He},d){}^{15}{\rm O}$
differential cross sections.
The squares are data points and the solid lines are the DWBA calculations
normalized to the experimental measurements in the main peaks; (a)- our 
data, (b) - our fit of the angular distribution measured in 
Ref. \protect\cite{champ2002} .}  
\label{fig_ang}
\end{figure}

The ${}^{14}$N + $p$ $\to$ ${}^{15}$O + $\gamma$ reaction
is a notorious example of an important astrophysical reaction
where the subthreshold state plays a dominant role. This reaction
is one of the most important processes in the CNO cycle.  As the
slowest reaction in the cycle, it defines the rate of energy
production \cite{rolfs88} and, hence, the lifetime of stars that
are governed by hydrogen burning via CNO processing. The
${}^{14}{\rm N}(p,\gamma){}^{15}{\rm O}$ reaction proceeds through
direct capture to the subthreshold state $3/2^{+}$, 6.79 MeV
(binding energy $504$ keV) and, possibly, via direct capture to
the ground state and resonant capture through the first
resonance and subthreshold resonance at $E_{s}=-504$ keV.
The overall normalization of the direct capture is defined by the
corresponding ANC. The ANC for the subthreshold state $E_{s}=-504$ keV also 
determines the partial proton width of the subthreshold resonance. 
In order to determine the ANCs for ${}^{14}{\rm N}+ p \to {}^{15}{\rm O}$,
the ${}^{14}{\rm N}({}^{3}{\rm He},d){}^{15}{\rm O}$ proton
transfer reaction has been measured at an incident energy of 26.3
MeV. Angular distributions for proton transfer to the ground and
five excited states were obtained. Angular distributions of
deuterons from the ${}^{14}$N(${}^{3}$He,$d$)${}^{15}$O reaction
leading to the  most important transition to the fourth excited
state $3/2^{+}$, 6.79 MeV in ${}^{15}{\rm O}$ measured by us 
at an incident energy of 26.3 MeV and in \cite{champ2002} 
measured at an incident energy of 20 MeV, together with our DWBA 
fits are shown in Fig.\@ \ref{fig_ang}.
The proton ANC that we obtain for the ${}^{14}{\rm N} +p \to
{}^{15}{\rm O}(3/2^{+},\,6.79$ MeV) is $C^{2}=27.1 \pm 6.8$
fm$^{-1}$. Using our ANCs, we calculated the
astrophysical factor and reaction rates for the ${}^{14}{\rm
N}(p,\gamma){}^{15}{\rm O}$ process.
\begin{figure}[tbp]
\includegraphics*[width=\linewidth]{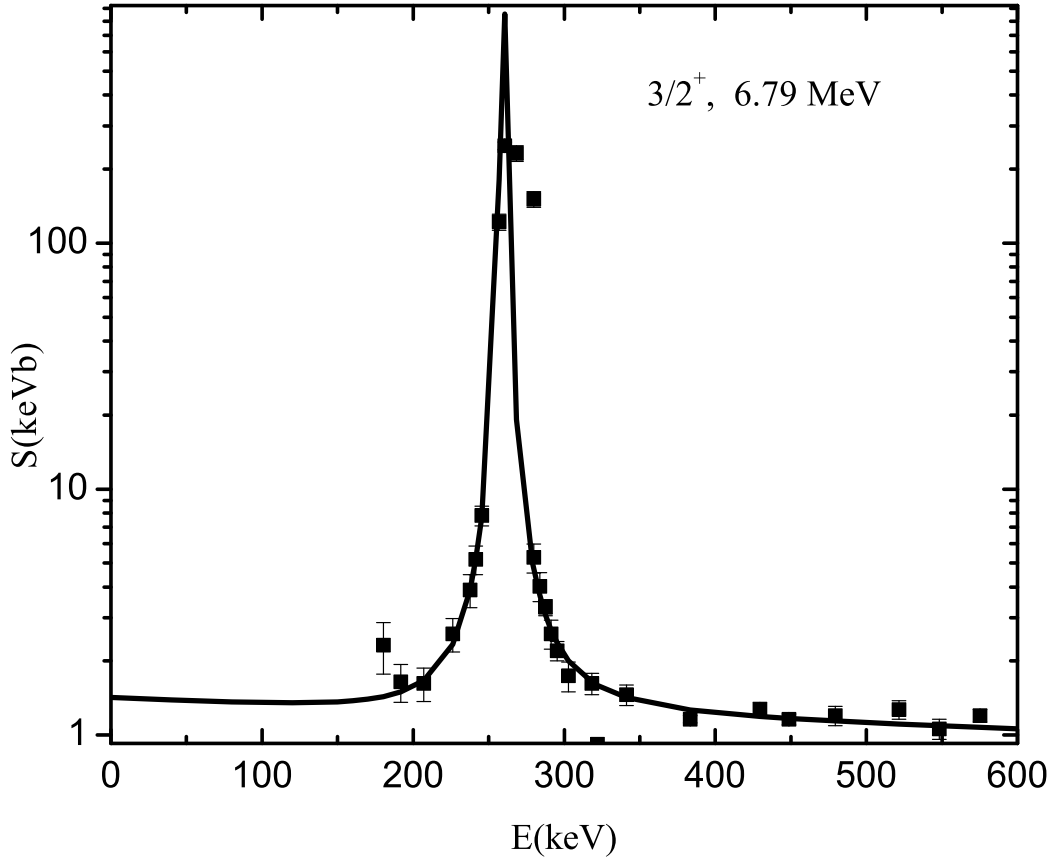}
\caption{The ${}^{14}{\rm N}(p, \gamma){}^{15}{\rm O}$
astrophysical $S$ factor for capture to the fourth
excited state ((c): $3/2^{+},\,6.79$ MeV), which includes the incoherent
sum of the resonant and nonresonant terms. The squares are data points 
\protect\cite{schroder87}; the solid lines represent the calculated
$S$ factor using our measured ANC. }  
\label{fig_sfctr}
\end{figure}
The capture to the $3/2^{+},\,6.79$ MeV state dominates all others and 
the calculated astrophysical factor is $S(0)= 1.40 \pm 0.20$ keV\,b.
The calculated and experimental $S(E)$ factors for
the transition to this subthreshold state are
presented in Fig.\@ \ref{fig_sfctr}.
The uncertainty in $S(0)$ is entirely determined by the ANC of this
state and the 13\% systematic uncertainty in the experimental $S(E)$ 
factor \cite{schroder87}. We find that the astrophysical factor for the 
capture to the ground state is $S(0)= 0.15 \pm 0.07$ keV\,b.
The total calculated astrophysical factor at zero
energy is $S(0)=1.70 \pm 0.22$ keV\,b what is in excellent agreement 
with the $S$ factor $S(0)=1.70 \pm 0.22$ keV\,b obtained from recent direct 
measurements performed at LUNA \cite{luna}. The lower astrophysical 
factor  of the ${}^{14}{\rm N}(p,\gamma){}^{15}{\rm O}$ reaction leads 
to an increase in the age of the main-sequence turnoff in globular 
clusters \cite{champ2001}.
\subsection{ANC and interference of direct and resonant amplitudes}
\label{ANCinterference}
To demonstrate how the information about the ANC can be used to 
determine the interference sign of the resonant and direct amplitudes of 
the radiative capture process we use the R matrix approach.
Let us consider the radiative capture reaction
$a + b \to c + \gamma$.

The $R$-matrix radiative capture amplitude to a state of nucleus
$c$ with a given spin $J_{f}$ and relative orbital angular momentum of 
the
bound state $l_f$ is given by the sum of resonant and nonresonant 
(direct capture) amplitudes \cite{bark91}:
\begin{equation}
U_{Il_f J_f J_i }=U_{Il_f J_f J_i }^R + U_{Il_f J_f J_i }^{NR},
\label{rmampl1}
\end{equation}
Interference effects only occur in Eq.\@ (\ref{rmampl1}) if the
resonant and nonresonant amplitudes have the same channel spin $I$ and
orbital angular momentum $l_{i}$.
In the one-level, one-channel approximation, the resonant
amplitude for the capture into the resonance with
energy $E_{rn}$ and spin $J_{i}$, and subsequent decay into the
bound state with the spin $J_{f}$, is given by
\begin{equation}
U_{I l_{i}J_{f}J_{i}}^{R}= -i\,e^{i(\phi_{l_{i}}- \sigma_{l_{i}})}\,
\frac{\lbrack\Gamma_{I 
l_{i}}^{J_{i}}(E)\rbrack^{1/2}\,\lbrack{\Gamma}_{\gamma
J_{f}}^{J_{i}}(E)\rbrack^{1/2}} {E-E_{rn}+i\,\frac{\Gamma_{J_{i}}}{2}}.
\label{resampl1}
\end{equation}
Here $J_{i}$ is the total angular momentum of the colliding nuclei
$a$ and $b$ in the initial state, $J_{a}$ and $J_{b}$ are the
spins of nuclei $a$ and $b$, and $I$, $k$, and $l_{i}$ are their
channel spin, relative momentum and orbital angular momentum in
the initial state. $U_{I l_{i}J_{f}J_{i}}$ is the transition
amplitude from the initial continuum state $(J_{i},\,I,\,l_{i})$
to the final bound state $(J_{f},\,I)$. Also
$\lbrack\Gamma_{Il_{i}}^{J_{i}}(E)\rbrack^{1/2}$ is real and its
square, $\Gamma_{Il_{i}}^{J_{i}}(E)$,  is the observable partial
width of the resonance in the channel $a+ b$ with the given set of
quantum numbers, $[{\Gamma}_{\gamma J_{f}}^{J_{i}}(E)]^{1/2}$ is
complex and its modulus square is the observable radiative width:
\begin{equation}
\Gamma_{\gamma\, J_{f}}^{J_{i}}(E) = |[{\Gamma}_{\gamma\, 
J_{f}}^{J_{i}}
(E)]^{1/2}|^{2}. \label{ggendep1}
\end{equation}
The energy dependence of the partial and radiative widths is given by
\begin{equation}
\Gamma_{I l_{i}}^{ J_{i}}(E)= \frac{P_{l_{i}}(E)}{P_{l_{i}}(E_{Rn})}\,
\Gamma_{I l_{i}}^{J_{i}}(E_{Rn}),  \label{prwdt1}
\end{equation}
and
\begin{equation}
\Gamma_{\gamma\,J_{f}}^{J_{i}}(E) =  (\frac{E + 
\varepsilon_{f}}{E_{Rn} +
\varepsilon_{f}})^{2\,L +1}\, \Gamma_{\gamma\,J_{f}}^{J_{i}}(E_{Rn}),
\label{gradwdt1}
\end{equation}
respectively.
Here, $\Gamma_{I l_{i}}^{J_{i}}(E_{Rn})$ and
$\Gamma_{\gamma\,J_{f}}^{J_{i}}(E_{Rn})$
are the experimental partial and radiative resonance
widths, $\varepsilon_{f}$ is the proton binding energy of the bound
state in nucleus $A$, $\,L$ is the multipolarity of the gamma quanta
emitted during the transition, and $\Gamma_{J_{i}} \approx\sum_{I}\,
\Gamma_{I l_{i}}^{J_{i}}$. In a strict $R$-matrix approach
\begin{equation}
\lbrack \Gamma_{\gamma\,J_{f}}^{J_{i}}(E)\rbrack^{1/2}  =
2\,\lbrack P_{l_{i}}(E) \rbrack^{1/2} \, \gamma_{\gamma J_{f}}^{J_{i}}.
\label{prtwdt2}
\end{equation}
Here the radiative reduced-width amplitude $\gamma_{\gamma 
J_{f}}^{J_{i}}$ is given by
the
sum of the internal and external (or channel) reduced-width
amplitudes:
\begin{equation}
\gamma_{\gamma J_{f}}^{J_{i}}= \gamma_{\gamma J_{f}}^{J_{i}}(int) +
\gamma_{\gamma J_{f}}^{J_{i}}(ch).  \label{redwdtamp1}
\end{equation}
Hence the total radiative width is
\begin{equation}
|\lbrack \Gamma_{\gamma\,J_{f}}^{J_{i}}(E)\rbrack|
=|\lbrack \Gamma_{\gamma\,J_{f}}^{J_{i}}(E)\rbrack^{1/2}_{int} +
\lbrack \Gamma_{\gamma\,J_{f}}^{J_{i}}(E)\rbrack^{1/2}_{ch}|^{2},
\label{Ggtotintch1}
\end{equation}
\begin{equation}
\lbrack \Gamma_{\gamma\,J_{f}}^{J_{i}}(E)\rbrack^{1/2}_{int,ch}  =
2\,\lbrack P_{l_{i}}(E) \rbrack^{1/2} \, \gamma_{\gamma 
J_{f}}^{J_{i}}(int,ch).
\label{ggwdt3}
\end{equation}
While the internal reduced-width amplitude is real, the channel
reduced-width
amplitude is complex \cite{bark91} and is defined as
\begin{eqnarray}
\gamma_{\gamma J_{f}}^{J_{i}}(ch)= i^{l_{i}+L-l_{f}+1}\,
e^{i(\omega_{l_{i}}-\phi_{l_{i}})}\,
\frac{1}{k}\,{\mu_{ab}}^{L + 1/2}\, \nonumber\\
\left( \frac{
Z_{a}\,e}{m_{a}^{L}}+
(-1)^{L}\,\frac{Z_{b}\,e}{m_{b}^{L}}\right)
\times  \sqrt{\frac{(L+1)(2\,L+1)}{L}}\, \nonumber\\
\frac{1}{(2\,L+1)!!}\,
(k_{\gamma}\,a)^{L+1/2}\,C_{J_{f}Il_{f}}\,
\sqrt{\Gamma_{b I l_{i}}^{J_{i}}(E_{R})} \nonumber\\
([F_{l_{i}}(k,a)]^{2} +
[G_{l_{i}}(k,a)]^{2})\,
\times W_{l_{f}}(2\,\kappa\,a)\,
(l_{i}0\, L0|l_{f}0)\,    \nonumber\\
U(L\, l_{f}\,J_{i}\,I;\,l_{i}\,J_{f})\,
J_{L}(l_{i}\,l_{f}).    \label{ggch1}
\end{eqnarray}
The nonresonant capture amplitude is given by
\begin{eqnarray}
U_{I l_{i}J_{f}J_{i}}^{NR}&=& -(2)^{3/2}\,i^{l_{i}+L-l_{f}+1}\,
e^{i(\omega_{l_{i}}-\phi_{l_{i}})}\,
\frac{1}{k}\,{\mu_{ab}}^{L + 1/2}\,  \nonumber\\
&&\times \left( \frac{%
Z_{a}\,e}{m_{a}^{L}}+
(-1)^{L}\,\frac{Z_{b}\,e}{m_{b}^{L}}\right) \,\sqrt{\frac{%
(L+1)(2\,L+1)}{L}}\, \nonumber \\
&&\times \frac{1}{(2\,L+1)!!}(k_{\gamma}\,a)^{L+1/2}\,C_{J_{f}Il_{f}}\,
F_{l_{i}}(k,r_{0})\, \nonumber\\
&& \times G_{l_{i}}(k,r_{0})\,W_{-\eta_{f},l_{f}+1/2}
(2\,\kappa\,r_{0})  \nonumber\\
&&\sqrt{P_{l_{i}}}(l_{i}0\, L0|l_{f}0)\,
U(L\, l_{f}\,J_{i}\,I;\,l_{i}\,J_{f})\, \nonumber\\
&& \times J^{^{\prime}}_{L}(l_{i}\,l_{f}), \label{nrsampl1}
\end{eqnarray}
\begin{equation}
P_{l_{i}}(E)= \frac{k\,r_{0}}{F_{l_{i}}^{2}(k,r_{0})+ 
G_{l_{i}}^{2}(k,r_{0})},
\label{pntr1}
\end{equation}
where $F_{l_{i}}$ and $G_{l_{i}}$ are the regular and singular (at the
origin) solutions of the radial Schr\"odinger
equation, $\kappa=\sqrt{2\mu_{ab}\,\varepsilon_{f}}$ is the wave
number, and $k_{\gamma}=E + \varepsilon_{f}$ is the
momentum of the emitted photon.
Integrals $J_{L}(l_{i}\,l_{f})$ and $J^{^{\prime}}_{L}(l_{i}\,l_{f})$ are 
expressed in terms of $F_{l_{i}},\, G_{l_{i}}$ and Whittaker function 
$W_{-\eta_{f},l_{f}+1/2}$ 
and are given in \cite{bark91,tang03}.
Both the channel radiative width and nonresonant amplitude are 
normalized in terms of the ANC, $C_{J_{f} I l_{f}}$, which defines the 
amplitude of the tail of the bound state wave function of nucleus $c$ 
projected onto the
two-body channel $a+b$ with the quantum numbers $J_{f},\,I,\,l_{f}$. 
Such a
normalization is physically transparent: both quantities describe  
peripheral processes and, hence, contain the tail of the overlap 
function of the bound wave functions of $c,\,a$ and $b$, whose 
normalization is given by the corresponding ANC. Note that in the 
$R$-matrix method the internal nonresonant amplitude is included into 
the resonance term.
Also, in the conventional $R$-matrix approach the channel radiative 
width and nonresonant amplitude are
normalized in terms of the reduced width amplitude, which is not
directly observable and depends on the channel radius. However, it is 
more convenient to
express the normalization of the nonresonant amplitude in terms of the
ANC that can be measured independently \cite{mukhtr99}.  Then only the
radial matrix element depends on the channel radius.
As we can see from Eqs (\ref{ggch1}) and (\ref{nrsampl1}) the relative 
phase of the channel radiative width and the nonresonant amplitude is 
fixed because only the ANC has unknown phase factor.
Thus by measuring the ANC for the bound state we are able to fix the 
absolute normalization of the channel radiative width and nonresonant 
amplitude simultaneously.
\subsection{Interference of the resonant and nonresonant amplitudes for
the ${}^{11}{\rm C}(p, \gamma){}^{12}{\rm N}$ astrophysical
radiative capture reaction} The evolution of very low-metallicity,
massive stars depends critically on the amount of CNO nuclei that
they produce. Alternative paths from the slow 3\,$\alpha$ process
to produce CNO seed nuclei could change their fate. The
${}^{11}{\rm C}(p, \gamma){}^{12}{\rm N}$ reaction is an important
branch point in one such alternative path. At energies appropriate
to
stellar evolution of very low-metallicity, massive stars, nonresonant 
capture to the ground state and interference of the second resonance and 
the nonresonant terms determine the reaction rate. The ANC for 
${}^{12}{\rm N} \to {}^{11}{\rm C} +p$ has been determined from 
peripheral transfer reaction ${}^{14}{\rm N}({}^{11}{\rm C},\,{}^{12}{\rm N})\\
{}^{13}{\rm C}$ at
10 MeV/nucleon \cite{tang03}. The contributions from the second
resonance and interference effects were estimated using the
R-matrix approach with the measured asymptotic normalization
coefficient and the latest value for the radiative width of the
second resonance \cite{riken02}. The ANC gives useful information
not only about the overall normalization of the direct capture
amplitude, but also about the radiative width of the resonances.
According to Eqs. (\ref{ggwdt3}), the channel part of the
radiative width may be determined from the ANC. Since the channel
part is complex, $\lbrack \Gamma_{\gamma
J_{f}\,J_{i}}(E)\rbrack^{1/2}_{ch} = \lambda+ i\tau$, while the
internal part of the radiative width amplitude is real, i. e.
$\lbrack \Gamma_{\gamma J_{f}\,J_{i}}(E)\rbrack^{1/2}_{int} =
\nu$, the total radiative width is given by
\begin{equation}
\Gamma_{\gamma J_{f}\,J_{i}}(E)= (\lambda+ \nu)^{2} + \tau^{2}.
\label{Ggtot1}
\end{equation}
The relative phase of $\lambda$ and $\nu$ is, a priori, unknown,
so these real parts may interfere either constructively or
destructively. Hence, $\tau^{2}$ always provides a lower limit for
the radiative width and additional stronger limits may be obtained
if assumptions are made about the interference between the two
real contributions. For constructive interference of the real
parts, the channel contribution gives a stronger lower limit. In
the case of destructive interference, if $|\lambda| > |\nu|$, the
channel contribution gives an upper limit for the radiative width.
These limits depend on only one model parameter, the channel
radius.

Recently, a measurement at RIKEN \cite{riken02} found the gamma 
width to be  $13.0 \pm 0.5$ meV. Using the measured ANC we 
find that for a channel radius of $r_{0}= 5.0$ fm, $\Gamma_{\gamma
J_{f}\,J_{i}}(E_{R})_{ch}= 54$ meV. Taking into account the
experimental value of the total radiative width, one can find the
internal contribution from 
\begin{equation}
\Gamma_{\gamma J_{f}\,J_{i}}(E_{R})=
|\Gamma_{\gamma J_{f}\,J_{i}}(E_{R})^{1/2}_{ch} +
\Gamma_{\gamma J_{f}\,J_{i}}(E_{R})^{1/2}_{int}|^{2}.
\label{Ggtotchint}
\end{equation}
There are two solutions, 15 and 112 meV. Assuming that the second
value is too high \cite{desc99}, we conclude that the internal
part of the radiative width is 15 meV, and destructive
interference between the real parts of the channel and internal
contributions gives the experimental value, 13 meV. In this case,
the channel contribution alone represents an upper limit for the
radiative width, while the square of the imaginary part of the
channel contribution, 1.8 meV, gives a lower limit. The relative
phase between the direct capture amplitude and the channel
contribution to the radiative width of the second resonance is
fixed in the R-matrix approach. Therefore, when the channel
contribution to the radiative width dominates, the sign of the
interference effects may be determined unambiguously. For
${}^{11}{\rm C}(p, \gamma){}^{12}{\rm N}$, we find that the
nonresonant and resonant capture amplitudes interfere
constructively below the resonance and destructively above it. It
has important consequences on the reaction rates for ${}^{12}{\rm
N}$ production. In particular, the reaction sequence $^7 {\rm Be}
(\alpha,\gamma)^{11}{\rm C}(p,\gamma)^{12}{\rm N}$ $^{7}{\rm
Be}(\alpha,\gamma){}^{11}{\rm C}(p, \gamma){}^{12}{\rm N}$ will
provide a means to produce CNO nuclei, while bypassing the 3
$\alpha$ reaction, in lower-density environments than previously 
anticipated \cite{wiescher89}.
\subsection{Interference of the resonant and nonresonant amplitudes for
the ${}^{13}{\rm N}(p, \gamma){}^{14}{\rm O}$ astrophysical radiative 
capture}
${}^{13}{\rm N}(p, \gamma){}^{14}{\rm O}$ is one of the key reactions 
which trigger the onset of the hot CNO cycle. This transition occurs 
when the proton capture rate on
${}^{13}{\rm N}$ is faster, due to increasing stellar temperature
$( \ge 10^8$ K), than the ${}^{13}{\rm N}$ $\;\beta$-decay rate. The 
rate of this reaction is dominated by the resonant capture to the ground 
state of ${}^{14}{\rm O}$  through the first excited state of $(E_R  = 
0.528$ MeV). However, through constructive interference, direct capture 
below the resonance makes
a non-negligible contribution to the reaction rate.
We have determined this direct contribution by measuring the
asymptotic normalization coefficient for ${}^{13}{\rm N} + p \to
{}^{14}{\rm O}(0.0$ MeV). This ANC  has been determined from the 
peripheral reaction
${}^{14}{\rm N}({}^{13}{\rm N}, {}^{14}{\rm O}){}^{13}{\rm C}$ 
\cite{tang04}.
The radiative capture cross section was estimated using an R-matrix 
approach with the measured asymptotic normalization coefficient and the 
latest resonance parameters.
What is not known is the sign of the interference term between the 
resonant and nonresonant components of the radiative capture amplitudes.
As we have mentioned it is possible to sometimes infer the sign of 
the interference to be used in an R-matrix calculations of the radiative
capture cross section if the ANC is known even in the absence of direct 
experimental data.
Such is the case for the reaction being considered here. At energies 
below the resonance, the channel part, which depends on the ANC, has the 
same sign as the nonresonant
amplitude leading to the constructive interference of these two terms.
From Eqs. (\ref{ggch1}) and (\ref{ggwdt3}) we find
${[\Gamma_{\gamma\,J_{f}}^{J_{i}}(E_{R})]}^{1/2}_{ch}= 0.90 + i\,0.02$ 
eV$^{1/2}$ and the channel radiative width
$|{[\Gamma_{\gamma\,J_{f}}^{J_{i}}(E_{R})]}_{ch}|= 0.81\times 10^{-6}$ 
eV at the resonance energy and the channel radius $r_{0}=5$ fm. The 
total resonance radiative width is
$|\lbrack \Gamma_{\gamma\,J_{f}}^{J_{i}}(E)\rbrack||= 
|[\Gamma_{\gamma\,J_{f}}^{J_{i}}(E)]^{1/2}_{int}
+ {{[\Gamma_{\gamma\,J_{f}}^{J_{i}}(E)]}^{1/2}_{ch}|}^{2}$. Thus there 
are two possible solutions for the internal part, a large negative value 
${[\Gamma_{\gamma\,J_{f}}^{J_{i}}(E)]}^{1/2}_{int{(1)}}= - 2.73$ 
eV$^{1/2}$ and a small positive value 
${[\Gamma_{\gamma\,J_{f}}^{J_{i}}(E)]}^{1/2}_{int{(2)}}= 0.93$ 
eV$^{1/2}$. The first solution leads to the destructive 
interference with the non-resonant component at energies below the resonance, 
but it yields a high internal radiative width, 
$|{\Gamma_{\gamma\,J_{f}}^{J_{i}}(E)]}_{int}|= 7.48$ eV. 
The second solution leads to the constructive interference 
with the non-resonant component at energies below the resonance peak. 
We select this second solution because it is corroborated by the microscopic 
calculations \cite{desbaye1989}, where it has been 
shown that the internal and external parts of the $E1$ matrix 
elements have the same sign and very close magnitudes.
Our choice is also supported by the single-particle
calculations \cite{descouvemont1999,tang04}.
Due to this constructive interference we find the S factor for 
${}^{13}{\rm N}(p, \gamma){}^{14}{\rm O}$ to be larger than previous 
estimates. Consequently, the transition from the cold to hot CNO cycle 
for novae would be controlled by the slowest proton capture reaction
${}^{14}{\rm N}(p, \gamma){}^{15}{\rm O}$.

\section{Trojan Horse}
The Trojan Horse method (THM) is a powerful indirect method which 
selects the quasi- free (QF) contribution of an appropriate three-body 
reaction performed at energies well above the Coulomb 
barrier to extract a charged particle two-body cross section at astrophysical 
energies free of Coulomb suppression. The THM has been suggested by Baur
\cite{baur86th} and has been advanced and practically applied by a group 
from the Universit{\'a} di Catania  working at the INFN-Laboratori
Nazionali del Sud in Catania in collaboration with
other Institutions (see \cite{spit04} and references
therein). The THM has already been applied many times to
reactions connected with fundamental astrophysical problems
\cite{copi95,piau02} such as ${}^{7}{\rm Li}(p,\alpha ){}^{4}{\rm He},
{}^{6}{\rm Li}(d,\alpha){}^{4}{\rm He},\\
{}^{6}{\rm Li}(p,\alpha ){}^3He$, and many others, 
see \cite{spit04} and references\\
therein. 

Let us consider the TH reaction
\begin{equation}
a+ A \to y+ b+ B, \label{THreaction1}
\end{equation}
where $a=(xy)$. The subreaction of interest is
\begin{equation}
x+A \to b+B. \label{subTHreaction1}
\end{equation}
In the TH method the incident particle $a$ is accelerated to
energies above the Coulomb barrier. After penetration through the
barrier the projectile breaks into $x+y$ leaving the
fragment $x$ to interact with target $A$, while the second
fragment-spectator $y$ leaves carrying away the excess energy.
By a proper choice of the final particle kinematics, the THM allows
one to extract the cross section of the sub-process
(\ref{subTHreaction1}). However the extracted amplitude of the
reaction (\ref{subTHreaction1}) in the THM is half-off-energy shell
because the initial particle $x$ in the sub-process
(\ref{subTHreaction1}) is off-the-energy shell. It has been
suggested in the original paper \cite{baur86th} that the
virtualiy of particle $x$ is compensated for by the higher 
momentum components in the Fermi motion of the fragments $x$ and $y$ 
inside the projectile $a$. However, high momentum components 
means that the distance between the fragments is so 
small that the interaction between the fragments is not 
negligible and the mechanism of the reaction is more 
complicated than the QF one. Instead, the virtuality of particle $x$
in the extracted cross section is significantly compensated if we take into 
account the binding energy of the fragments $x$ and $y$ in the projectile $a$
\cite{tumino03}.

\begin{figure}[tbp]
\includegraphics*[width=\linewidth]{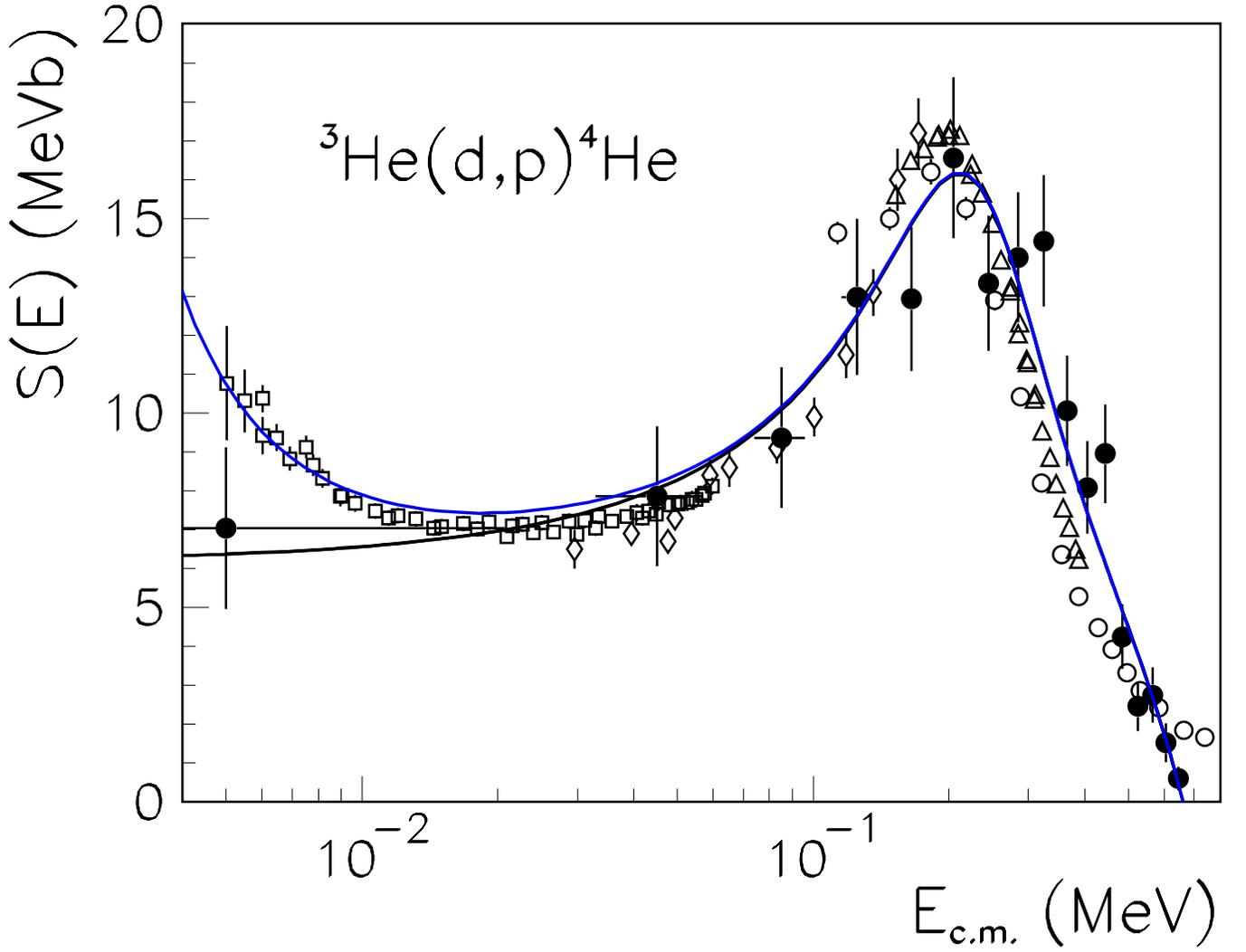}
\caption{The ${}^{3}{\rm He}(d,p){}^{4}{\rm He}$ astrophysical $S$
factor determined from the TH reaction. The open circles and open
triangles are direct experimental data; the full dots are the TH
data. The black solid line represents the behavior of the bare
nucleus S(E)-factor, resulting from a fit on the TH data, while
the solid blue line is interpolation of the direct data.}  
\label{fig_TH}
\end{figure}

The THM allows one to determine both direct and resonant reactions
(\ref{subTHreaction1}). As an example of the result achieved using
the THM, we present in Fig. \ref{fig_TH} the astrophysical factor
for the ${}^{3}{\rm He}(d,p){}^{4}{\rm He}$ process determined from the
${}^{3}{\rm He}({}^{6}{\rm Li},\alpha\,p){}^{4}{\rm He}$ TH
reaction \cite{lacognata04}. The TH resonant cross section (full
dots) is normalized to the direct experimental data (open circles
and open triangles) at energies near the resonance peak. The black
solid line is the result of a fit of the TH data (see
ref.\cite{lacognata04} for details), showing the trend of the bare
nucleus S(E)-factor, while the blue solid line is obtained by
interpolating the screened direct data.

\subsection{TH reaction amplitude}
\label{THreactampl}

\begin{figure}[tbp]
\includegraphics*[width=\linewidth]{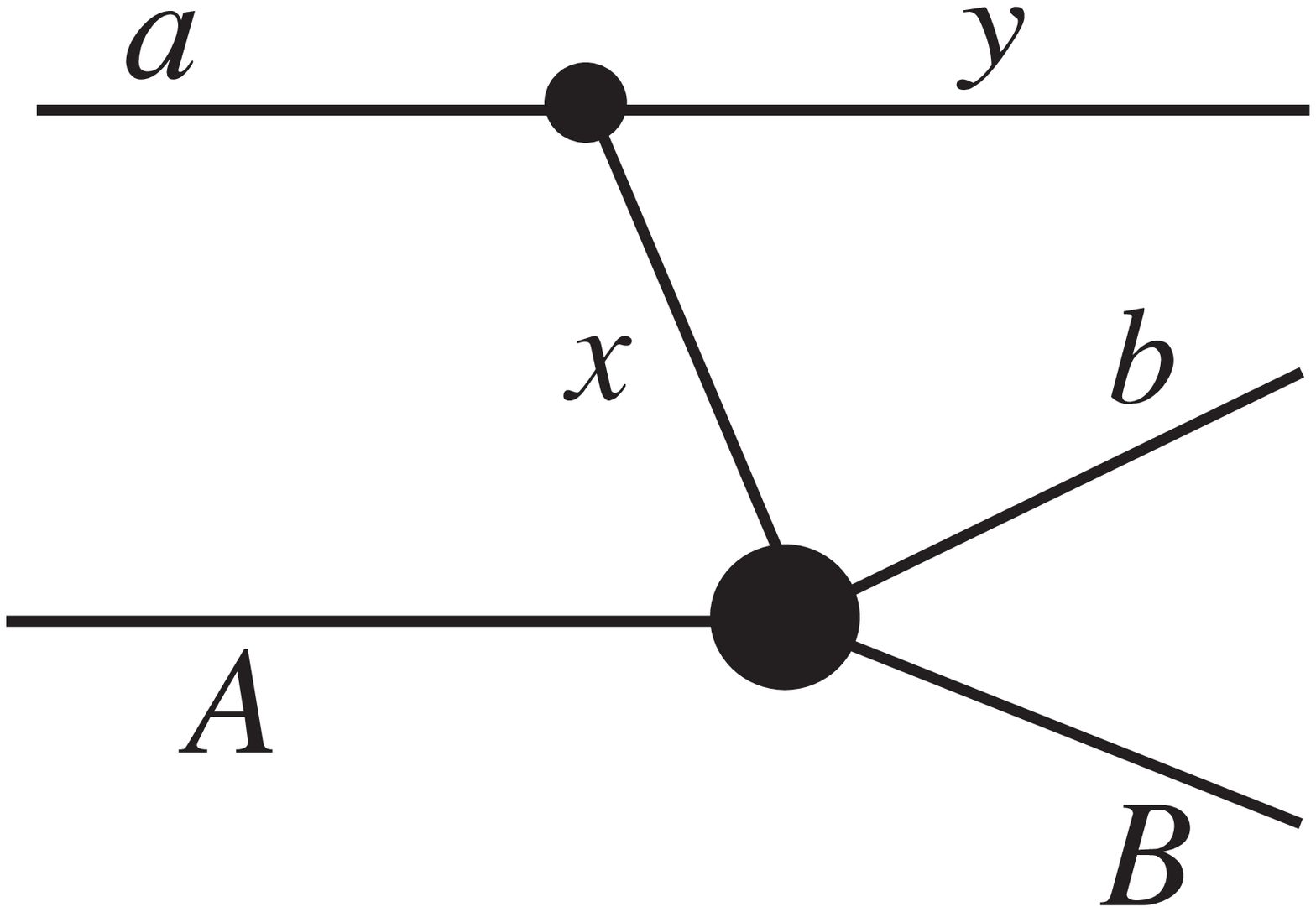}
\caption{Pole diagram describing the quasi-free mechanism.}  
\label{Polediagram}
\end{figure}

A simple mechanism describing the TH process is the so-called QF process 
shown in Fig. \ref{Polediagram}.
In the quasi-free process it is assumed that the incident particle 
(assume incident particle is $A$) interacts with one of the fragments
of $a=(xy)$, say with $x$ which is considered to 
be "quasifree", while the second fragment is considered to be a "passive" spectator 
which is not involved in the process. Thus the interaction of the spectator 
$y$ with $x$ and $A$ in the knockout process is neglected . The fact that the fragment $x$ is not free is taken into account by folding the quasi-free reaction amplitude with the 
Fourier component of the $(xy)$ bound-state wave function which takes into account the Fermi motion of $x$ in the bound state $a=(xy)$. 
  
In this section we present a derivation of the TH reaction 
amplitude from the general $2 \to 3$ reaction 
amplitude for the TH process (\ref{THreaction1}).
A general expression for the amplitude of the reaction is given by
\begin{eqnarray}
M =  < \chi _{bB}^{( - )} \chi _{yF}^{( - )}\varphi _y \varphi _b \varphi _B |\Delta V_f (1 + G^{+}\Delta V_i )|\varphi _A \varphi _a 
\chi _i^{( + )}> \nonumber\\
 \label{THpost1} \\
 =  < \chi _{bB}^{( - )} \chi _{yF}^{( - )} \varphi _y \varphi _b \varphi _B |(\Delta V_f\, G^{+} + 1)\Delta V_i |\varphi _A \varphi _a \chi _i^{( + )}>.
\nonumber\\ 
\label{THprior1}
\end{eqnarray}
The amplitudes (\ref{THpost1}) and (\ref{THprior1}) are the post and  
prior forms of the exact amplitude. Let us consider the post form.
Here, $G^{+}$ is the total Green function of the system $a + A$, $\,\chi^{(+)}_{i}$ is the distorted wave describing the scattering 
wave function of $a+A$ in the initial state of the reaction, $\chi^{(-)}_{bB}$ is the distorted wave describing the scattering of particles $b+ B$ in the final state: the distorted wave $\chi^{-}_{yF}$ describes the distorted
wave of the spectator $y$ and the center of mass of the system $F=b+B$ in the final state. 
For the moment we assume that Coulomb interactions are screened. Eventually we can take the limit of the screening radius to infinity. Also $\phi_{i}$ is the bound state wave 
function of nucleus $i$, 
\begin{eqnarray}
\Delta V_{i} = V_{aA} - U_{aA},     \label{deltavi1} \\
\Delta V_{f}= V_{bB} -U_{bB} + V_{yF} - U_{yF},  \label{deltavf1}
\end{eqnarray}              
$V_{ij}$ and $U_{ij}$ are the interaction potential and optical potential 
between particles $i$ and $j$.  For example, $V_{aA}= V_{xA} + V_{yA}$. 
To extract the amplitude of the subprocess $x+ A \to b+B$, which is the final goal of the TH method, we note that the Hamiltonian of the system
$a+A$ is 
\begin{equation}
H= H_{aA} + H_{a}+H_{A}= H_{xA} + H_{yF} + H_{x} + H_{A} + H_{y}, 
\label{hamiltonian1} 
\end{equation}
where $H_{i}$ is the internal Hamiltonian of nucleus $i$ and $H_{ij}=
T_{ij} + V_{ij}$ is the Hamiltonian of the relative motion of nuclei $i$ 
and $j$, $T_{ij}$ is their relative kinetic energy operator and $V_{ij}$ is their interaction potential. 
The total Green's function operator can be written as 
\begin{eqnarray}
G^{+}&=&\frac{1}{E - H_{aA} - H_{a} - H_{A}+i0} 
\label{totgrfnct1} \\
&&= \frac{1}{E - H_{xA} - H_{yF}-H_{xyA}^{0}+i0}\nonumber\\
\label{totgrfnct2} \\
&&= \frac{1}{E - H_{xA} - T_{yF}-U_{yF} - \Delta{V}_{yF}-H_{xyA}^{0}+i0}\nonumber\\
\label{totgrfnct3}\\
&&= {\tilde G}^{+} + G^{+}\,\Delta V_{yF}\,{\tilde G}^{+},
\label{greenfunct1}
\end{eqnarray}
Here $\Delta V_{f}= V_{yF}- U_{yF}$, $\,V_{yF}= V_{yx} + V_{yA}$,\\
$H_{xyA}^{0}=H_{x}+H_{y}+ H_{A}$ and
\begin{equation}
{\tilde G}^{+}=\frac{1}{E - H_{xA} - T_{yF}- U_{yF} - H_{xyA}^{0}+i0} \label{tildgrfnct1}
\end{equation}
We substitute Eq. (\ref{tildgrfnct1}) into (\ref{THpost1}) and drop 
the term \\
$\Delta V_{f}\,G^{+}\,\Delta V_{yF}\,{\tilde G}^{+}\,\Delta V_{i}$ 
as the higher order term in the perturbation expansion over $\Delta V$.
Then we get from Eq. (\ref{THpost1})
\begin{eqnarray}
M =  < \chi _{bB}^{( - )} \chi _{yF}^{( - )}\varphi _y \varphi _b \varphi _B |\Delta V_f (1 + {\tilde G}^{+}\,\Delta V_i )|\varphi _A \varphi _a 
\chi _i^{( + )}>. \nonumber\\
 \label{THpost12} 
\end{eqnarray}
To single out the TH subprocess amplitude we replace $\Delta V_{i}=
V_{xA} + V_{yA} - U_{aA}$ by $V_{xA}$ and $\Delta V_{f}= V_{yF} - U_{yF}
+ V_{bB} - U_{bB}$ by $\Delta V_{bB}=V_{bB} - U_{bB}$. Then the amplitude 
(\ref{THpost1}) becomes
\begin{eqnarray}
M =  < \chi _{bB}^{( - )} \chi _{yF}^{( - )}\varphi _y \varphi _b \varphi _B |\Delta V_{bB}\,(1 + {\tilde G}^{+}\,V_{xA} )|\varphi _A \varphi _a 
\chi _i^{( + )}> \nonumber\\
=  < \chi _{bB}^{( - )} \chi _{yF}^{( - )}\varphi _y \varphi _b \varphi _B |\Delta V_{bB}\,(1 + G^{+}_{xA}\,V_{xA} )|\varphi _A \varphi_a 
\chi _i^{( + )}>.  \nonumber\\
\label{THposti1} 
\end{eqnarray}
Here
\begin{equation}
G^{+}_{xA}=\frac{1}{E_{xA} - H_{xA} +i0} \label{grfnctxA1}
\end{equation}
and $E_{xA}$ is the relative kinetic energy of particles $x$ and $A$.
The appearance of $G^{+}_{xA}$ in Eq. (\ref{grfnctxA1}) is due to 
\begin{eqnarray}
<\chi _{yF}^{( - )}\varphi _y \varphi _b \varphi _B|{\tilde G}^{+} =
<\chi _{yF}^{( - )}\varphi _y \varphi _b \varphi _B|G^{+}_{xA}. 
\label{gtildegxa1}
\end{eqnarray} 
Eq. (\ref{THposti1}) reveals a very important result.
It contains a factor $1 + G^{+}_{xA}\,V_{xA}$. For the on-shell case,
the relative momentum of particles $x$ and $A\;{\rm {\bf p}}_{xA}= 
{\rm {\bf k}}_{xA}$, where ${\rm {\bf k}}_{xA}$ is the $x-A$ on-shell relative momentum related with their relative kinetic energy as  
$E_{xA}= p_{xA}^{2}/(2\,\mu_{xA})$.  Correspondingly, 
\begin{eqnarray}
(1 + G^{+}V_{xA} )|e^{i{\rm {\bf k}}_{xA}  \cdot {\rm {\bf r}}_{xA} }>  = \chi_{{\rm {\bf k}}_{xA} }^{+}({\rm {\bf r}}_{xA}). 
\label{onshwf1}
\end{eqnarray}
is the scattering wave function of particles $x$ and $A$ interacting via the optical potential $V_{xA}$. We assume at the moment that all the Coulomb interactions are screened. However, in the TH method the entry particle $x$ is not free because it is in the bound state $a=(xy)$, i. e. the momentum of $x$ is not fixed. In other words, $x$ is off-the-energy shell because $E_{xA} \not= p_{xA}^{2}/(2\,\mu_{xA})$.   
For the off-shell case  
\begin{eqnarray}
(1 + G^ +  V_{xA} )|e^{i{\rm {\bf p}}_{xA}  \cdot {\rm {\bf r}}_{xA} } > = \chi _{(os) {\rm {\bf k}}_{xA}, {\rm {\bf p}}_{xA} }^{+}({\rm {\bf r}}_{xA} )
\label{offshwf1}
\end{eqnarray}
is the so-called off-shell scattering function,

\subsection{TH method for direct reactions}

\begin{figure}[tbp]
\includegraphics*[width=\linewidth]{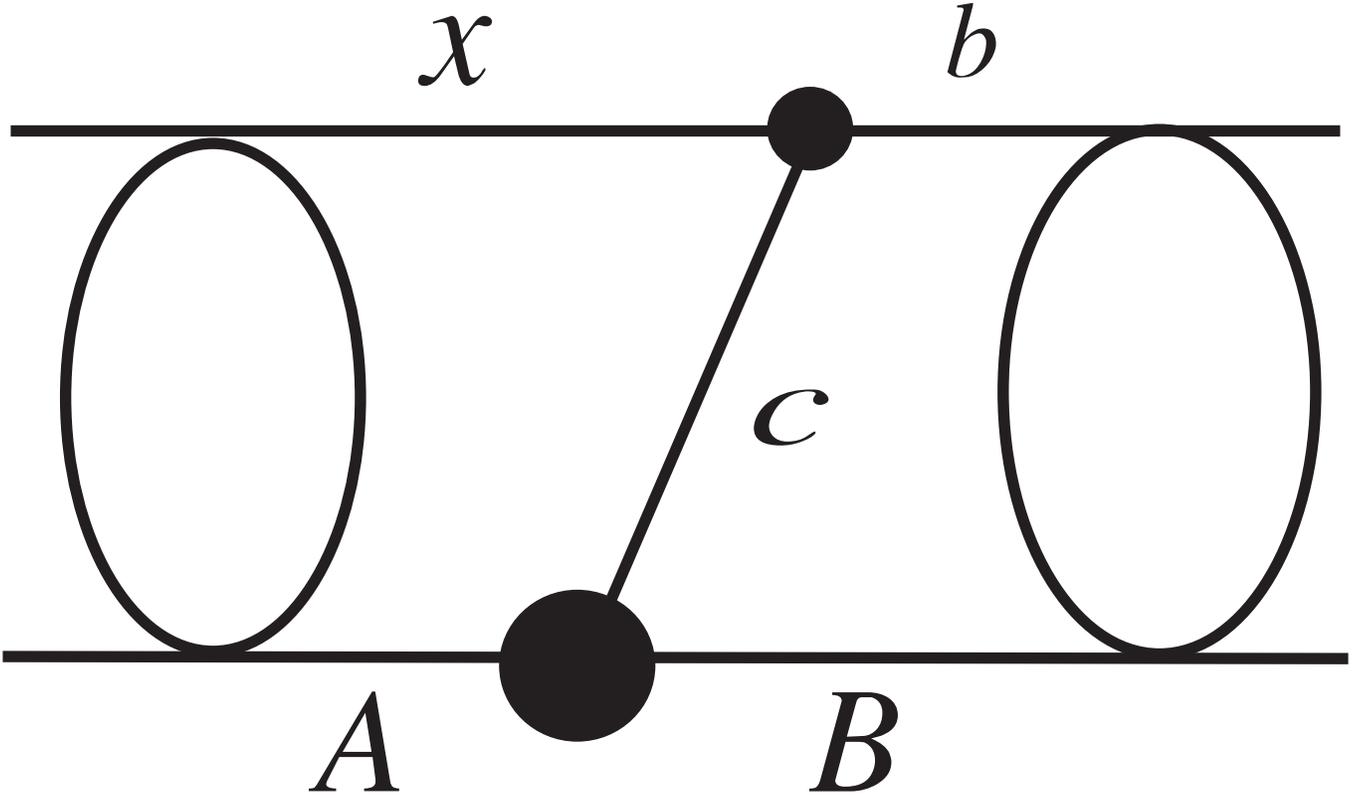}
\caption{Pole diagram describing the direct reaction $x + A \to b +B$ mechanism. Bubbles show the initial and final state interactions.}  
\label{THdirectlast}
\end{figure}

We first consider the direct subreaction (\ref{subTHreaction1}).
We assume that this reaction proceeds through the transfer of particle
$c$ from $A$ to $x$ (it can be also considered as a particle transfer from $x$ to $A$), i. e. $A=(Bc)$ and $b=(xc)$. The "pole" diagram corresponding to the on-shell reaction describing the particle $c$ transfer mechanism with the $x-A$ rescattering in the initial and $b-B$ rescattering in the final state is shown in Fig. \ref{THdirectlast}. This 
diagram describes the DWBA amplitude. 
To simplify Eq. (\ref{THposti1}) in the case of the direct transfer subprocess, we insert the projection operators $\sum \varphi_{x}><\varphi_{x}$,
$\,\sum \varphi_{B}><\varphi_{B}$ and $\sum \varphi_{c}><\varphi_{c}$ into the bra and ket states. The sum is taken over discrete states and an integral is used for the continuum states of the corresponding nucleus. We leave in the projection operator only the ground state projections $\varphi_{x}><\varphi_{x}$, $\,\varphi_{B}><\varphi_{B}$ and $\varphi_{c}><\varphi_{c}$ assuming that  only the ground states of $x$, $B$ and $c$ contribute to the reaction. 
If necessary the excited states can also be taken into account.
Then we get 
\begin{eqnarray}
<\varphi _y \varphi _b \varphi _B |\Delta V_{bB}\,(1 + G^{+}_{xA}\,V_{xA} )|\varphi _A \varphi_a> \nonumber\\
\approx  <\varphi _b|\varphi_{c}\varphi_{x}> <\varphi _B |\Delta V_{bB}|\varphi_{B}> \nonumber\\
\times (1 +
<\varphi_{x}| G^{+}_{xA}|\varphi_{x}> <\varphi_{x}| V_{xA}|\varphi_{x}>)
\nonumber\\
<\varphi_{c}\varphi_{B}|\varphi_A> <\varphi_{x}\varphi _y |\varphi_a> 
\label{projector1}
\end{eqnarray}
We introduce the overlap functions $I^{\alpha}_{\beta \gamma}= <\varphi_{\beta}\,\varphi_{\gamma}|\varphi_{\alpha}>$ and use the approximation 
$\,<\varphi_{x}| V_{xA}|\varphi_{x}>\,\approx U_{xA}$;
also we use the approximation  
\begin{eqnarray}
<\varphi_{x}| G^{+}_{xA}|\varphi_{x}> \approx G^{(U)+}_{xA}=
{(E_{xA} - T_{xA} - U_{xA} +i0)}^{-1}. \nonumber\\ 
\label{guapprox1}
\end{eqnarray}
Note that
$<\varphi _B |\Delta V_{bB}|\varphi_{B}> \approx V_{xB} + V_{cB} - U_{bB}$,
where $V_{jB}$ is the interaction potential between the point like nuclei $j=x,c$ and $B$. All the neglected terms are higher order in the perturbation theory over $\Delta V$. 
Then we get in lowest order for the TH amplitude with the subprocess described by the direct transfer reaction (\ref{subTHreaction1}): 
\begin{eqnarray}
M = < \chi _{yF}^{( - )}[ \chi _{bB}^{( - )} I^{b}_{xc}|\Delta V_{bB}|\,
I^{A}_{cB}\,(1 + G^{+}_{xA}\,V_{xA} )]\, I^{a}_{xy}\chi _i^{( + )}>. \nonumber\\
\label{MTHdirect1}
\end{eqnarray}
The expression in the brackets is the amplitude of subreaction 
(\ref{subTHreaction1}) which is the final goal of the TH. To see it we just
rewrite (\ref{MTHdirect1}) in momentum space:
\begin{eqnarray}
M &=& \int \frac{ {\rm d}{\rm {\bf p}}_{yF} }{(2\,\pi)^{3}}\,\frac{{\rm d}{\rm {\bf p}}_{xA}}{(2\,\pi)^{3}}\,  \chi _{yF}^{*( - )}({\rm {\bf p}}_{yF})\,
M^{sub}({\rm {\bf k}}_{bB},{\rm {\bf p}}_{xA})\, 
I^{a}_{xy}({\rm {\bf p}}_{xy})\,\nonumber\\
&&\times \chi _i^{( + )}({\rm {\bf p}}_{xA}), 
\label{MTHdirect2}
\end{eqnarray}
where
\begin{equation}
{\rm {\bf p}}_{xy}= \frac{m_{y}{\rm {\bf p}}_{x} - m_{x}{\rm {\bf p}}_{y}}
{m_{x}+m_{y}}
= \frac{m_{y}}{m_{x}}{\rm {\bf p}}_{a} - {\rm {\bf p}}_{y}.
\label{pxy1}
\end{equation}
Also note that in the center of mass of TH reaction 
$a + A \to y+b+B$ the relative momentum is given  by
${\rm {\bf p}}_{aA}= {\rm {\bf p}}_{a}$ and 
${\rm {\bf p}}_{yF}= {\rm {\bf p}}_{y}$. We denote by ${\rm {\bf p}}_{i}$
$\;({\rm {\bf k}}_{i})$ the momentum of the virtual (real) particle $i$ and by
$\;({\rm {\bf k}}_{ij})$ the relative momentum of virtual (real) particles $i$ and $j$. Also
$\chi _i^{( + )}({\rm {\bf p}}_{xA}) \equiv  \chi _{{\rm {\bf k}}_{aA}}^{( + )}({\rm {\bf p}}_{xA})$, i. e. it is the Fourier component of the $a-A$ scattering wave function with the incident momentum ${\rm {\bf k}}_{aA}$ which in the center of mass of the TH reaction is just ${\rm {\bf k}}_{a}$. Correspondingly
$\chi _{yF}^{( - )}({\rm {\bf p}}_{yF}) \equiv \chi _{{\rm {\bf k}}_{yF}}^{( - )}({\rm {\bf p}}_{yF})$.

\begin{figure}[tbp]
\includegraphics*[width=\linewidth]{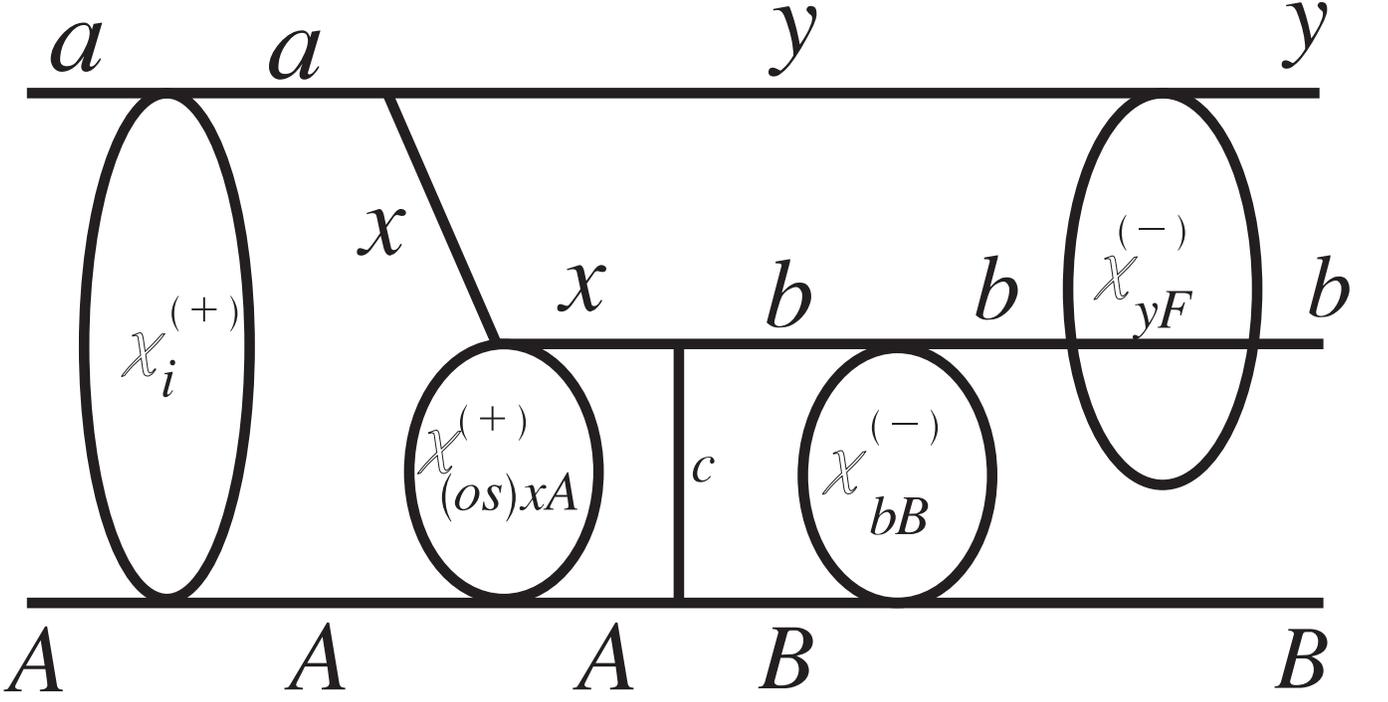}
\caption{Diagram describing the TH reaction $a + A \to y+ b+B$ proceeding through the direct subprocess  $x + A \to b +B$ mechanism. Bubbles show initial and final state interactions and the off-shell scattering function.}  
\label{fig_THFakh}
\end{figure}

The half-off-the-energy shell amplitude of the subprocess 
(\ref{subTHreaction1}) is given by 
\begin{eqnarray}
M^{sub}({\rm {\bf k}}_{bB},{\rm {\bf p}}_{xA})= < \chi _{bB}^{( - )} I^{b}_{xc}|\Delta V_{bB}|\,
I^{A}_{cB}\,\chi _{(os) {\rm {\bf k}}_{xA}, {\rm {\bf p}}_{xA} }^{+}>.  \nonumber\\
\label{hlfoffshsubpramp1}
\end{eqnarray}
The virtuality of the entry particle $x$ of this amplitude results in the fact that the relative momentum of particles $x$ and $A$ in the initial state of reaction 
(\ref{subTHreaction1}) $ p_{xA} \not= \sqrt{2\,\mu_{xA}\,E_{xA}}$.
Due to the off-shell entry particles amplitude (\ref{hlfoffshsubpramp1})
does not have the Gamow penetration factor.
We would like to underscore that from $E_{xA} + Q= E_{bB}$ for positive 
$Q >0$ for reaction (\ref{subTHreaction1}) at $E_{xA} \to 0$, $\,E_{bB} \approx {\rm const}$.
Hence the off-shell scattering function 
$\chi _{(os) {\rm {\bf k}}_{xA}, {\rm {\bf p}}_{xA} }^{+}$ is the only $E_{xA}$  dependent factor in $M^{sub}$ at $E_{xA} \to 0$. The off-shell scattering function is 
a universal factor which does not depend on the specifics of the direct reaction. 
Rewritting matrix element in Eq. (\ref{hlfoffshsubpramp1}) in the momentum representation gives
\begin{eqnarray}
M^{sub}({\rm {\bf k}}_{bB},{\rm {\bf p}}_{xA})= \int \frac{ {\rm d}{\rm { \bf p}}_{bB} }{(2\,\pi)^{3}}\,\frac{ {\rm d}{\rm { \bf p}}^{'}_{xA} }{(2\,\pi)^{3}}
\chi _{  {\rm { \bf k}}_{bB} }^{*(-)}({\rm { \bf p}}_{bB}) \nonumber\\
\times I^{*b}_{xt}
({\rm { \bf p}}^{'}_{x} - \frac{m_{x}}{m_{b}}\,{\rm { \bf p}}_{b})\,\Delta V_{bB}\,
I^{A}_{cB}({\rm { \bf p}}_{B} - \frac{m_{B}}{m_{A}}\,{\rm { \bf p}}^{'}_{A}) \nonumber\\
\times \chi _{(os) {\rm {\bf k}}_{xA}, {\rm {\bf p}}_{xA} }^{+}({\rm { \bf p}}^{'}_{xA})
\label{hlfoffshsubpramp2}
\end{eqnarray}
Approximation $\Delta V_{bB}\approx V_{cB}$, which works for $m_{x} > m_{c}$, is enough for us to investigate the dependence  of \\ 
$M^{sub} ({\rm {\bf k}}_{bB},{\rm {\bf p}}_{xA})$
on $E_{xA}$ for arbitrary masses of $x$ and $c$. Using this approximation we get 
from Eq. (\ref{hlfoffshsubpramp2})
\begin{eqnarray}
M^{sub}({\rm {\bf k}}_{bB},{\rm {\bf p}}_{xA})= \int \frac{ {\rm d}{\rm { \bf p}}_{bB} }{(2\,\pi)^{3}}\,\frac{ {\rm d}{\rm { \bf p}}^{'}_{xA} }{(2\,\pi)^{3}}
\chi _{  {\rm { \bf k}}_{bB} }^{*(-)}({\rm { \bf p}}_{bB}) \nonumber\\
\times I^{*b}_{xc}
({\rm { \bf p}}^{'}_{x} - \frac{m_{x}}{m_{b}}\,{\rm { \bf p}}_{b})\,
W^{A}_{cB}({\rm { \bf p}}_{B} - \frac{m_{B}}{m_{A}}\,{\rm { \bf p}}^{'}_{A}) \nonumber\\
\times \chi _{(os) {\rm {\bf k}}_{xA}, {\rm {\bf p}}_{xA} }^{+}({\rm { \bf p}}^{'}_{xA}).
\label{hlfoffshsubpramp3}
\end{eqnarray}
Here $W^{A}_{cB}({\rm { \bf p}}_{cB})$ is the form factor determined by
\begin{equation}
W^{A}_{cB}({\rm { \bf p}}_{cB}) = \int {\rm d}{\rm { \bf r}}_{cB} 
e^{-i\,{\rm { \bf p}}_{cB} \cdot {\rm { \bf r}}_{cB} }\, V_{cB}(r_{cB})\,
I^{A}_{cB}({\rm { \bf r}}_{cB}).
\label{formfactor1}
\end{equation}
The  Fourier component of the off-shell scattering function 
$\chi _{(os) {\rm {\bf k}}_{xA}, {\rm {\bf p}}_{xA} }^{+}({\rm { \bf r}}_{xA})$
is given by
\begin{eqnarray}
\chi _{(os) {\rm {\bf k}}_{xA}, {\rm {\bf p}}_{xA} }^{+}({\rm { \bf p}}^{'}_{xA})
&=& \delta({\rm { \bf p}}^{'}_{xA} - {\rm { \bf p}}_{xA}) + 
G^{+}_{0}( p^{'}_{xA}; E_{xA}) \nonumber\\
&&\times T({\rm { \bf p}}^{'}_{xA},{\rm { \bf p}}_{xA}; E_{xA}), \label{offshscst11}
\end{eqnarray}
\begin{equation}
G^{+}_{0}( p^{'}_{xA}; E_{xA}) =
\frac{1}{E_{xA}- p_{xA}^{' 2}/2\,\mu_{xA} +i0},
\label{freegrfnct1}
\end{equation}
$T({\rm { \bf p}}^{'}_{xA},{\rm { \bf p}}_{xA}; E_{xA})$ is the off-shell $x-A$ 
scattering amplitude. 
Amplitude $M^{sub}({\rm {\bf k}}_{bB},{\rm {\bf p}}_{xA})$ extracted from the 
THM should be compared with the corresponding on-shell reaction amplitude
\begin{eqnarray}
M^{onsh}({\rm {\bf k}}_{bB},{\rm {\bf k}}_{xA})= \int \frac{ {\rm d}{\rm { \bf p}}_{bB} }{(2\,\pi)^{3}}\,\frac{ {\rm d}{\rm { \bf p}}_{xA} }{(2\,\pi)^{3}}
\chi _{ {\rm { \bf k}}_{bB} }^{*(-)}({\rm { \bf p}}_{bB}) \nonumber\\
\times I^{*b}_{xt}
({\rm { \bf p}}_{x} - \frac{m_{x}}{m_{b}}\,{\rm { \bf p}}_{b})\,\Delta V_{bB}\,
I^{A}_{cB}({\rm { \bf p}}_{B} - \frac{m_{B}}{m_{A}}\,{\rm { \bf p}}_{A}) \nonumber\\
\times \chi_{{\rm {\bf k}}_{xA}}({\rm { \bf p}}_{xA})
\label{hlfoffshsubpramp3}
\end{eqnarray} 
Eqs. (\ref{MTHdirect2}) and (\ref{hlfoffshsubpramp1}) is our final result. The diagram corresponding to this amplitude  (\ref{MTHdirect2}) is shown in Fig. \ref{fig_THFakh}.
Eq. (\ref{MTHdirect2}) is a general expression for the TH reaction amplitude which contains the half-off-shell direct subprocess amplitude and the initial and final state 
rescatterings. As we can see the subprocess amplitude is not factorized, but instead is folded with the initial and final state distorted waves and the overlap function for $a \to y + x$. Note that if the initial and distorted waves in the momentum space are replaced by delta-functions, Eq. (\ref{MTHdirect2}) just becomes a trivial plane wave impulse approximation described by the diagram of Fig. \ref{fig_TH}.

\subsection{TH for resonant reactions}
In Subsection \ref{THreactampl} we derived a general expression, Eq. (\ref{THposti1}),
for the amplitude of the TH reaction (\ref{THreaction1}) which is valid for both direct and resonant subprocesses (\ref{subTHreaction1}).  
Here we consider the resonant TH reactions, i. e. we assume that the 
subprocess
(\ref{subTHreaction1}) proceeds through the intermediate resonance 
$F^{*}$. 
Our goal is to relate the half-off-shell and on-shell resonant 
amplitudes. Note that it is easier to relate the off-shell and 
on-shell resonant reactions than the direct ones.
The resonant TH amplitude can be extracted from Eq. (\ref{THposti1}) in a straightforward manner because it contains the Green's operator $G^{+}_{xA}$.  Below we demonstrate 
how to do it. For simplicity here we neglect the initial and final state interactions.

\begin{figure}[tbp]
\includegraphics*[width=\linewidth]{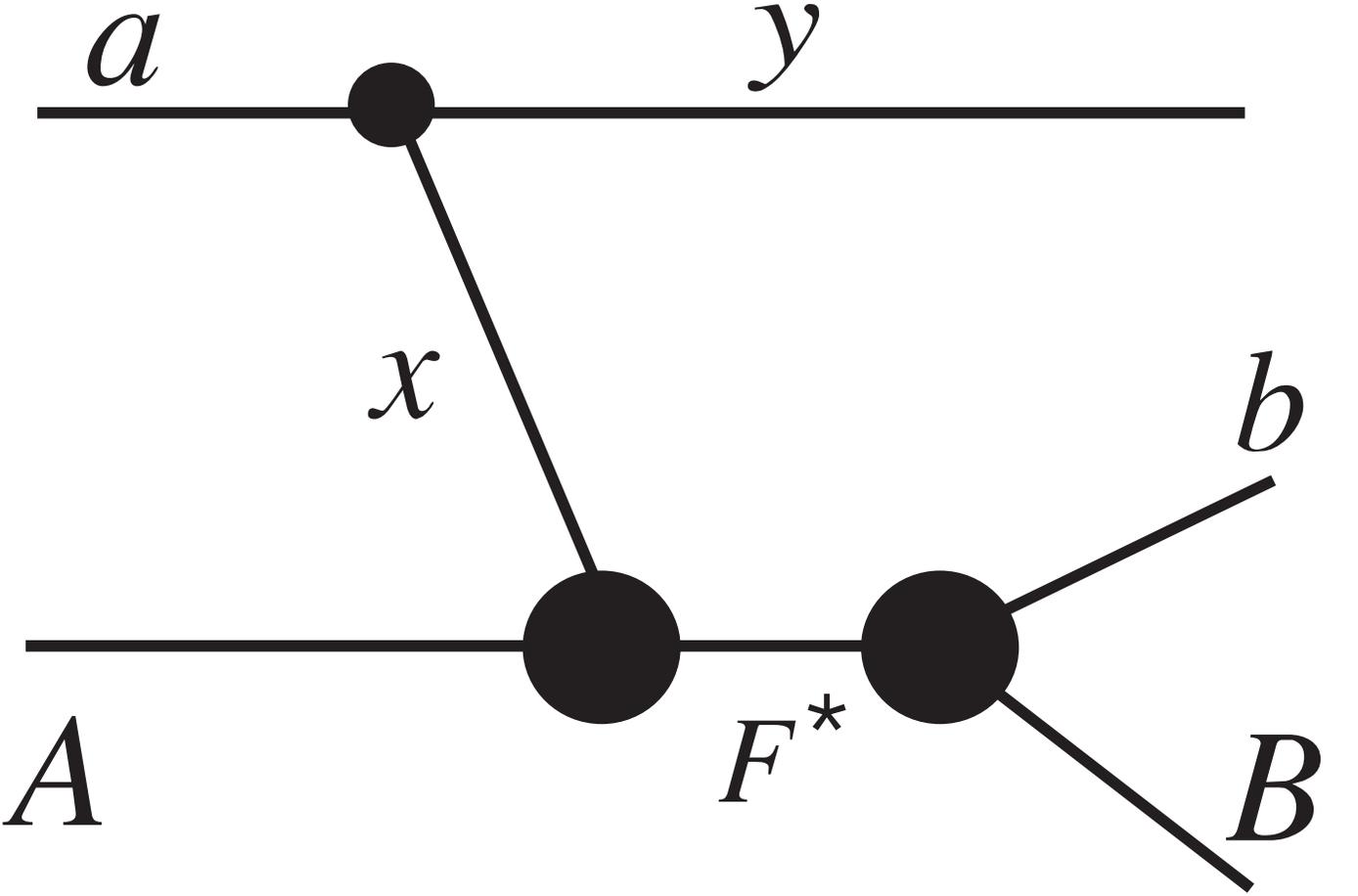}
\caption{Diagram describing the resonant reaction $a + A \to y+ b+B$.}  
\label{fig_THresonance}
\end{figure}

The TH amplitude of the
reaction (\ref{THreaction1}), which proceeds through the resonance state 
$F^{*}$ in the intermediate system $x+ A$, is given by
\begin{equation}
M = M^{sub(R)}({\rm {\bf k}}_{bB},{\rm {\bf p}}_{xA})\, 
I^{a}_{xy}({\rm {\bf p}}_{xy}). 
\label{THampl1}
\end{equation}
Here $M^{sub(R)}$ is the amplitude of the resonant subprocess 
(\ref{subTHreaction1}). 
Usually in practical calculations the overlap function $I^{a}_{xy}$ is expressed in terms of the corresponding single-particle bound state wave function $\varphi_{xy}$:
\begin{equation}
I^{a}_{xy}= S_{xy}^{1/2}\,\varphi_{xy}. \label{bstwf1}
\end{equation}
Here, $S_{xy}$ is the spectroscopic factor of the bound state $(xy)$ in $a$ with given quantum numbers. For simplicity we don't write down symbols corresponding to the quantum numbers and assume that $S_{xy}=1$.  
In the momentum space the bound state wave function is given by
\begin{equation}
\varphi _{xy} ({\rm {\bf p}}_{xy} )\, =  - 2\mu _{xy} \frac{{W({\rm 
{\bf p}}_{xy} )\,}}{{p_{xy}^2 \, + \,\kappa_{xy}^2 }},
\label{bndstwf1}
\end{equation}
\begin{eqnarray}
 W({\rm {\bf p}}_{xy} ) = \,\int {d{\rm {\bf r}}} \,e^{ - i{\rm {\bf 
p}}_{xy}  \cdot {\rm {\bf r}}} \,V_{xy} (r)\,\varphi _{xy}({\rm {\bf 
r}}) \\
  = ( - \varepsilon _{a}  - \frac{{p_{xy}^2 }}{{2\mu_{xy} 
}}\,)\,\varphi _{xy}
({\rm {\bf p}}_{xy} ),
\label{vertfrmfctr1}
\end{eqnarray}
Now we can find the virtuality factor
\begin{equation}
\sigma _x  = E_x  - \frac{{p_x^2 }}{{2m_x }}.
\label{vrtfact1}
\end{equation}
of the virtual particle $x$ using the energy and momentum
conservation laws in both vertices $a \to x+ y$ and $x+ A \to
F^{*}$. After simple algebraic transformations we get
\begin{eqnarray}
\sigma _x  =  E_{xA}  - \frac{{p_{xA}^2 }}{{2\mu _{xA} }}
=  - \frac{1}{{2\mu _{xy} }}[p_{xy}^2  + (\kappa _{xy}^a )^2 ]\, < 0.
\label{virtfact2}
\end{eqnarray}
Thus we derived a very important result for the relative kinetic energy of particles 
$x$ and $A$ $\,E_{xA}$ in the TH method: $\;E_{xA}  < p_{xA}^2 /2\mu _{xA}$, i. e. always $k_{xA} < p_{xA}$, where $k_{xA}= \sqrt{2\,\mu_{xA}\,E_{xA}}$ is the $x-A$ relative on-shell momentum.  The half-off-shell resonant
reaction amplitude in the TH method is described by the diagram shown in Fig.
\ref{fig_THresonance}
and is given by
\begin{eqnarray}
 M^{sub(R)} ({\rm {\bf k}}_{bB} ,\,{\rm {\bf p}}_{xA} ; E) =  - 
\frac{1}{2}(4\pi )^2 \,\sqrt {\frac{1}{{\mu _{bB} k_{bB} }}}  \nonumber\\
\sum\limits_{m_0  = -l_0 }^{l_0 } {Y_{l_0 m_0 } } (\hat{\rm {\bf k}}_{bB} 
Y_{l_0 m_0 }^* (\hat{\rm {\bf p}}_{xA} )\,e^{i\delta _{fl_0 }(k_{bB} 
)} \nonumber\\
\frac{{\sqrt {\Gamma _{bB} (E_{bB} ,r_0 )} \,w_{l_0 } (p_{xA} 
,k_{xA(R)} )}}{{E_{xA}  - E_{xA}^{(R)} }}. \nonumber\\
\label{resreact11}
\end{eqnarray}
Here $k_{xA(R)} = \sqrt{2\,\mu_{xA}\,E_{xA}^{(R)}}$, $\,{\rm {\bf
k}}_{bB}$ is the on-shell relative momentum of particles $b$ and $B$
in the final state, $l_{0}(m_{0})$ is the resonance orbital
angular momentum (its projection), $Y_{l_0 m_0 }$ is the
corresponding spherical harmonics, $\hat{\rm {\bf r}}= {\rm {\bf
r}}/r$, $\,\delta _{fl_0 }$ is the nonresonant (potential)
scattering phase shift of particles $b$ and $B$ in the final
state. The off-shell form factor
\begin{eqnarray}
 w_{l_0 } (p_{xA},k_{xA(R)}) = \int\limits_0^\infty  {dr\,r^2 \psi _{nl_0 
}^{(R)} (r)} \,V(r)j_{l_0 } (p_{xA}\,r)
\nonumber\\
= (E_{xA}^{(R)}  - E_{p_{xA}})\int\limits_0^\infty  {dr\,r^2 \psi _{nl_0 }^{(R)} 
(r)} \,j_{l_0 } (p_{xA}\,r) \nonumber\\
= (E_{xA}^{(R)}  - E_{p_{xA}} )\,\psi _{nl_0 }^{(R)}(p_{xA}).\nonumber\\
\label{offshfrmfactor1}
\end{eqnarray}
Here $\psi _{nl_0 }^{(R)}(r)$ is the resonant Gamow radial wave
function, $\psi _{nl_0 }^{(R)}(p_{xA})$ is its Fourier component,
$j_{l_0 } (p_{xA}\,r)$ is the spherical Bessel function,
$E_{p_{xA}}=p_{xA}^{2}/2\,\mu_{xA}$, $n$ is the principal quantum number. 

Let us write down the well known expression for the on-shell Breit-Wigner
resonance amplitude for the resonant process $x + A \to b + B$
\begin{eqnarray}
M^{(R)} ({\rm {\bf k}}_{bB},\,{\rm {\bf k}}_{xA}; E)\, = - 
\frac{1}{4}(4\pi )^2 \,\sqrt {\frac{1}{{\mu _{bB} k_{bB} }}}
\sqrt {\frac{1}{{\mu _{xA} k_{xA}}}} \nonumber\\
\times \sum\limits_{m_0  = l_0 }^{l_0 } {Y_{l_0 m_0 } } (\hat{\rm {\bf 
k}}_{bB} ) Y_{l_0 m_0 }^* (\hat{\rm {\bf k}}_{xA} )\,e^{i\delta _{fl_0 
}(k_{bB} )} e^{i\delta _{fl_0 }(k_{xA} )} \nonumber\\
\times \frac{{\sqrt {\Gamma _{bB} (E_{bB} ,r_0 )}\,\sqrt{\Gamma _{bB} 
(E_{xA} ,r_0 )} }}{{E_{xA}  - E_{xA}^{(R)} }}, \nonumber\\
\label{onshresampl1}
\end{eqnarray}
where ${\rm {\bf k}}_{xA}$ is the on-shell relative momentum of
the initial particles $x$ and $A$ and ${\rm {\bf k}}_{bB}$ is the
on-shell relative momentum of the final particles $b$ and $B$.
In the $R$-matrix method the resonance width contains the Coulomb-centrifugal 
barrier penetrability factor which exponentially decreases with energy. 
Hence for $E_{xA} \to 0$ the resonant amplitude $M^{R} \sim \sqrt{P_{l_{0}}(k_{xA})}\,{\tilde M}^{R}$. Just this factor makes it difficult or impossible to measure resonant 
reactions at astrophysically relevant energies. Now we compare the half-off-shell resonant 
amplitude, Eq. (\ref{resreact11}), and the on-shell amplitude, Eq. (\ref{onshresampl1}).
The half-off-shell amplitude contains the form factor \\
$w_{l_0 }(p_{xA},k_{(xA)R} )$. 
The barrier factor should come from the integral 
representation in Eq. (\ref{offshfrmfactor1}), namely from $j_{l_{0}}(p_{xA}r)$.
However, $j_{l_{0}}(p_{xA}r)$  does not contain the Coulomb penetration factor and does not depend on the on-shell momentum $k_{xA}$. Hence in limit $k_{xA} \to 0$ the off-shell form factor does not go to zero. We underscore that it is very important that always in the TH reaction $p_{xA} > k_{xA}$.  
Comparing Eqs (\ref{resreact11}) and (\ref{onshresampl1}) we get
\begin{eqnarray}
M^{(R)} ({\rm {\bf k}}_{bB} ,\,{\rm {\bf k}}_{xA} ; E_{xA} )\, =  
- \frac{1}{2}e^{i\delta 
_{(xA)l_0 } (k_{xA} )}
\sqrt {\frac{1}{{\mu _{xA} k_{xA} }}} \nonumber\\
\times \frac{{\sqrt {\Gamma _{xA} (E_{xA} ,r_0 )} }}{{w_{l_0 } (p_{xA} 
,k_{xA(R)} )}}
M^{sub(R)} ({\rm {\bf k}}_{bB} ,\,{\rm {\bf p}}_{xA} ; E_{xA} ).
\label{offonresamp1}
\end{eqnarray}
Note the only difference between the half-off-shell and the
on-shell resonant amplitudes is the appearance of the form factor
$w_{l_0}(p_{xA} ,k_{xA(R)})$. Now we give the expression for the
on-shell resonant cross section which can be derived from the TH
half-off-shell resonant cross section
\begin{eqnarray}
 \sigma (E_{xA} ) = \,\frac{{\mu _{xA} k_{xA} \,\mu _{bB} k_{bB} 
}}{{(2\pi )^2 }}\,\frac{1}{{k_{xA}^2 }}\frac{1}{{4\pi }}\int 
{{\rm d} \Omega _{{\rm {\bf k}}_{bB} }} \nonumber\\
 \times \int {{\rm d} \Omega _{{\rm {\bf k}}_{xA} }} |M^{(R)} ({\rm {\bf k}}_{bB} ,\,{\rm {\bf k}}_{xA} ; E_{xA} 
)\,\,|^2  \\
= \frac{1}{4}\frac{{\,\mu _{bB} k_{bB} }}{{(2\pi )^2 
}}\frac{1}{{k_{xA}^2 }}\frac{1}{{4\pi }}\frac{{\Gamma _{xA} (E_{xA} ,r_0 
)}}{{|w_{l_0 }^{} (p_{xA} ,k_{xA(R)} )|^2 }} \nonumber\\
\times \int {{\rm d} \Omega _{ {\rm {\bf k}}_{bB}} \,} \int 
{{\rm d} \Omega _{{\rm {\bf p}}_{xA} }} 
|M^{sub(R)} ({\rm {\bf k}}_{bB} ,\,{\rm {\bf p}}_{xA} ; E_{xA} )\,|^2.
\label{onofshcrsect1}
\end{eqnarray}

\section{Summary}
In this work we have addressed two important indirect techniques in nuclear astrophysics
use, asymptotic normalization coefficient (ANC) and Trojan Horse (TH) \\
method. Both techniques allow one to determine the astrophysical factors at Gamow peak or even at zero energy avoiding extrapolation procedure. 
The ANC method determines the overall normalization of the peripheral radiative capture processes. The ANC technique becomes especially powerful for astrophysical processes
proceeding through a subthreshold state - a loosely bound state. In this case the ANC determines both the overall normalization of the direct radiative capture to the subthreshold state and the resonance partial width for captures through the subthreshold resonance. We demonstrated the application of the ANC technique for the key CNO cycle reaction ${}^{14}{\rm N}(p,\,\gamma){}^{15}{\rm O}$. The ANC method turns out to be useful 
also for determination of the sign of the interference term of the resonant and nonresonant radiative capture amplitudes. We demonstrated it for two important CNO cycle reactions: ${}^{11}{\rm C}(p,\,\gamma){}^{12}{\rm N}$ and ${}^{13}{\rm N}(p,\,\gamma){}^{14}{\rm O}$. 

The TH method allows one to determine the astrophysical factors for astrophysical reactions, both direct and resonant. In practical applications the astrophysical factor
extracted from the TH reaction is available in a wide energy range from astrophysical energies to higher energies. Its absolute normalization is determined by normalization of the TH astrophysical factor to the one obtained from direct measurements at higher energies. Assuming that the energy dependence of the TH astrophysical factor is correct, one can determine the absolute astrophysical factor at astrophysical energies. In this work we have derived a general expression for the TH reaction amplitude which takes into account the off-shell effects and initial and final state interactions. The direct and resonant TH reactions are considered separately. We derived the TH amplitude for direct subreactions in terms of the off-shell scattering wave function. The energy dependence of this wave function determines the energy dependence of the TH astrophysical factor for an arbitrary direct reaction mechanism. We connect the TH resonant cross section with the 
on-shell resonant cross section. We intend to use the derived equations to calculate the absolute astrophysical factors. 

\begin{acknowledgments}
This work was supported by the U.\,S. 
Department of Energy under Grant No.\@ DE-FG03-93ER40773, the U.\,S. National 
Science Foundation under Grant No.\@ INT-9909787 and Grant No. \@ PHY-0140343, 
ME 385(2000) and ME 643(2003) projects NSF and MSMT, CR, project K1048102 and grant No. 202/05/0302 of the Grant Agency of the Czech Republic, and by the Robert A. Welch Foundation
\end{acknowledgments}

\end{document}